\documentclass[epj]{svjour}
\usepackage{amsmath} 
\usepackage{hyperref} 
\usepackage{graphicx}
\usepackage{array}
\usepackage[usenames,dvipsnames,svgnames,table]{xcolor}
\usepackage{bm}
\usepackage{subcaption}
\usepackage{float}
\usepackage{caption}
\usepackage[square,numbers]{natbib}
\usepackage{hypernat}
\begin{document}
\title{Three-dimensional {turbulence} effects on 
{plankton} dynamics behind an obstacle}
\author{Alice Jaccod\inst{1} \and Stefano Berti\inst{2} \and Enrico Calzavarini\inst{2} \and Sergio Chibbaro\inst{1} \\
}                     
%
%
\institute{Sorbonne Universit\'e, CNRS, UMR 7190, Institut Jean Le Rond d'Alembert, F-75005 Paris, France \and Univ. Lille, ULR 7512 - Unit\'e de M\'ecanique de Lille Joseph Boussinesq (UML), F-59000 Lille, France}
\date{Received: date / Revised version: date}
%
\abstract{
{We study a predator-prey model of plankton dynamics in the two and three-dimensional wakes of turbulent flows behind a cylinder, focusing on the impact of the three-dimensional character of the carrying velocity field on population variance spectra and spatial distributions. By means of direct numerical simulations, we find that the qualitative behavior of the biological dynamics is mostly independent of the space dimensionality, which suggests that only the relation between the typical flow and biological timescales is crucial to observe persistent blooms. Similarly, in both cases, we find that the spectral properties  of the planktonic populations are essentially indistinguishable from those of an inert tracer.  
The main difference arising from the comparison of the two and three-dimensional configurations concerns the local spatial distribution of plankton density fields. Indeed, the three-dimensional turbulent dynamics tend to destroy the localized coherent structures characterizing the two-dimensional flow, in which the planktonic species are mostly concentrated, thus reducing the phytoplankton biomass in the system.}
\PACS{
      {PACS-key}{discribing text of that key}   \and
      {PACS-key}{discribing text of that key}
     } 
} 
\maketitle
%


\section{Introduction}
\label{intro}
Phytoplankton produces organic matter in the surface ocean from sunlight through  
photosynthesis, thus providing the source of energy for almost all marine living creatures, including 
zooplankton, fish and larger animals~\cite{ML2005}. The efficiency of this 
process depends on several factors, such as the rate of essential biochemical reactions, 
the effectiveness of predation by higher organisms, nutrient abundance, and light availability. 
All these factors are 
affected by the transport and mixing processes taking place in the fluid environment 
where phytoplankton organisms live.

Understanding the relations between laminar and turbulent flows,  
and the distributions of planktonic species is a complex problem that has been 
previously addressed from different perspectives~\cite{abra,denman1995biological,martin2002,martin2003phytoplankton,lo,GR2008,GF2020}.
{Particular interest has been put on the characterization of the statistical features of plankton density fields, in terms of variance fluctuation spectra~\cite{DP1976,Smith_etal_1988,LK2004,franks}. Within this framework, a major question is whether the scaling of planktonic spectra is different from that of an inert quantity that is passively transported by a turbulent flow. The issue is relevant both at a fundamental level, to assess the relative importance of fluid and reactive (biological, in the present case) dynamics, and to quantify the patchiness (meaning, heterogeneity) of plankton spatial distributions. Indeed, were the spectra of planktonic fields different from those of a passive (non-reactive) scalar in a given wavenumber (or frequency) range, this would indicate the dominance of biological activity in the corresponding interval of spatial (or temporal) scales. Moreover, the slope of such spectra gives information on the scale-by-scale intensity of plankton fluctuations and, hence, could allow to quantify the typical size of structures where the biological species are possibly more concentrated.}

{Relying on dimensional arguments, some single-species models~\cite{DP1976,DP1976b} of plankton dynamics in three-dimensional (3D) turbulence predict that, at sufficiently small scales, the population fluctuation spectrum should scale as that of the velocity field ({i.e. according to the Obhukov-Corrsin scaling} as is the case for a non-reactive scalar). Due to the biological dynamics, however, the planktonic spectrum should flatten at larger scales, which would correspond to reduced patchiness in this range of scales. An extension of this picture was obtained by considering two plankton populations interacting according to Lotka-Volterra dynamics~\cite{powell}. Such a model reproduces the results of the single-species model 
when the interactions are neglected. In the presence of interactions, instead, it suggests that the reactive contribution to the spectrum of the density fluctuations of each species should be steeper than the spectrum of a non-reactive scalar (which would correspond to increased patchiness). Nevertheless, the full planktonic spectra result to be the sum of such a contribution and a non-reactive one, with different weights, making it difficult to draw general conclusions. In two-dimensional (2D) turbulent flows, forced at large scale, the same dimensional reasoning predicts flat spectra of reactive species (flatter than those of velocity fluctuations). 
Both for a single population and two interacting ones, the spectral slope is the same as the one found for the concentration of an inert substance.}

{Several observational studies report about the comparison of 
plankton density fluctuation spectra, obtained through the measurement of fluorescence (a proxy for phytoplankton concentration), with theoretical predictions (see~\cite{franks} for a critical review in 3D turbulent flows). The results are overall varied over different oceanic regions and do not appear conclusive~\cite{LK2004}. In some cases, quite flat spectra, possibly suggesting the importance of biological activity, have been found (see, e.g., \cite{DEROT2015210}). Recent detailed numerical simulations of multicomponent reactions, of different orders, in fully developed homogeneous isotropic 3D turbulence, instead, have shown that reactive spectra are indistinguishable from those of non-reactive passive scalar fields~\cite{wu}.}

In this work, we aim at comparing the 2D and 3D advection-reaction-diffusion dynamics of a predator-prey system, corresponding to the phytoplankton-zooplankton (PZ) model, in the turbulent flow generated by a cylindrical obstacle.  
Our motivation is twofold. On one side we are interested in testing the robustness of results already found in the 2D case~\cite{jaccod2021predator}, about the minimal flow ingredients needed to sustain a persistent bloom, when adding an extra dimension. On the other, based on the remarkably different cascade processes of 2D and 3D turbulence~\cite{alexakis2018cascades}, we intend to evaluate the impact of the ensuing multiscale turbulent dynamics, in two and three spatial dimensions, on the statistical features of the plankton population densities.
It is also worth noting that the majority of early studies addressing the conditions for plankton blooms under flow, which have been instructive to elucidate the basic mechanisms controlling the interplay between fluid and reactive dynamics, adopted the paradigm of chaotic advection, relying on 2D kinematic (i.e. synthetic) flows~\cite{toroczkai1998advection,neu,NLHP2002,fer,hl}. While it might appear, to some extent, reasonable that the overall phenomenology remains unchanged in the presence of 2D dynamic turbulent velocity fields, as it has been indeed recently shown~\cite{jaccod2021predator}, this is not straightforward when considering the extension to 3D authentic turbulence.  

The model investigated here, clearly, cannot be considered as fully representative of real plankton dynamics in the ocean. Our obstacle represents an idealized island (or obstruction of other kind), and we do not account for either vertical boundaries or topographic effects. Moreover, by considering PZ population dynamics, we neglect possible nutrient heterogeneities. While this choice allows us to limit the complexity of the problem, it precludes the possibility to describe oligotrophic environments, where nutrients are a growth limiting factor.
{Nevertheless, in our view, beyond the interest for general aspects related to the effect of turbulence on reaction dynamics, the present approach can be seen as a preliminary step towards the investigation of more realistic configurations relevant to plankton dynamics. In particular, it may reveal useful for subsequent developments aimed at improving the, still incomplete, understanding of the role of vertical transport (of different species) on different scales, which is considered an essential factor for primary production~\cite{levy2001impact,levy2003mesoscale,Levy_etal_2018}.}

This article is organized as follows. In Sec.~\ref{sec:math}, we introduce the model system and its governing equations. Section~\ref{sec:num3d} illustrates the numerical setup, as well as the flow configuration and the main parameters used. The results and their analysis are presented in Sec.~\ref{sec:results}. Section~\ref{sec:conclu3d} summarizes the main findings and discusses their implications.


\section{{Model dynamics}}
\label{sec:math}
We consider two plankton species interacting as in a predator-prey system, the phytoplankton (prey) and the zooplankton (predator). 
Their spatiotemporal evolution in a fluid flow can be conveniently described using coupled advection-reaction-diffusion equations. 
Following~\cite{jaccod2021predator}, we adopt for the reaction kinetics the PZ model~\cite{tru}.
By introducing a characteristic (obstacle) length $l_0$, 
a typical fluid velocity $u_0$, a typical time $l_0/u_0$, and the {phytoplankton} carrying capacity $K$, 
the nondimensional evolution equations for the population densities of phytoplankton $P(\boldsymbol{x},t)$ and of zooplankton $Z(\boldsymbol{x},t)$ [with $\boldsymbol{x} = (x,y,z)$] are:
\begin{subequations}
\begin{align}
& \partial_t P +  \bm{u} \cdot \bm{\nabla} P - \frac{1}{Re Sc}\bm{\nabla}^2 P = \left(\beta P\left(1-P\right) - \delta Z \frac{P^2}{P^2 + \chi^2}\right),\label{apza}\\
& \partial_t Z +  \bm{u} \cdot \bm{\nabla} Z - \frac{1}{Re Sc}\bm{\nabla}^2 Z =  \gamma Z \left( \delta \frac{P^2}{P^2 + \chi^2} - \lambda \right), \label{apzb}
\end{align} 
\end{subequations}
where $Re = u_0 d/\nu$ is the Reynolds number based on the obstacle diameter $d = 2l_0$, $\nu$ the viscosity coefficient, $Sc = \nu/D$ the Schmidt number, and $D$ the diffusion coefficient. 
The remaining biological parameters are $\beta = rl_0/u_0$, where $r$ is the maximum specific growth rate of $P$, $\delta = R_m l_0/u_0$, where $R_m$ is the maximum specific predation rate of $Z$,  $\chi = \kappa/K$, where $\kappa$ indicates how quickly that maximum is attained, 
while $\lambda = \mu l_0/(u_0 \gamma)$, with $\mu$ the rate of zooplankton removal  {(due to death or sinking)} and $\gamma$ the ratio of biomass consumed to biomass of new herbivores produced. 

{The space-independent version of the above model (representative of a well-mixed situation) can display excitability, and for this reason it was originally introduced to describe algal blooms~\cite{tru}. 
It should be noted that, for this to occur, it is necessary that two different timescales exist: to escape the predation control, and thus initiate the outbreak, the phytoplankton growth rate must be larger than the predation rate by the zooplankton.
We also recall that the space-independent reactive dynamics are characterized by three fixed points~\cite{tru}: $(P_1,Z_1)=(0,0)$, which represents the extinction of both species; $(P_2,Z_2)=(1,0)$, which gives the equilibrium value of $P$ in the absence of $Z$; and $(P_3,Z_3)=(P_{eq}, Z_{eq})$ where $P_{eq}=  \chi\sqrt{\lambda/(\delta-\lambda)}$ and $Z_{eq}= \beta(1-P_{eq})(P_{eq}^2+\chi^2)/(P_{eq}\delta)$, which is a stable equilibrium point for the parameter values here adopted.}

{For the velocity field $\bm{u}(\bm{x},t)$ appearing in Eqs.~(\ref{apza}) and (\ref{apzb}), we consider an incompressible 3D flow 
in the presence of a circular cylinder of diameter $d$ and height 
$L \gg d$, which is the solution of the Navier-Stokes equation with the appropriate boundary conditions (see Sec.~\ref{sec:num3d}).} 
{In nondimensional form, the Navier-Stokes equation, supplemented by the incompressibility condition, reads:
\begin{subequations}
\begin{eqnarray}
\partial_t \bm{u} + (\bm{u} \cdot \bm{\nabla}) \bm{u} & = & -\bm{\nabla} p + \frac{1}{Re} \bm{\nabla}^2 \bm{u}, \label{ansa}\\
\bm{\nabla} \cdot  \bm{u}  & = & 0, 
\label{ansb}
\end{eqnarray}
\end{subequations}
where 
$\bm{u}=(u_x,u_y,u_z)$ 
and $p$ is pressure. For the analogous 2D dynamics, the cylinder reduces to a circle and $u_z=0$.}


\section{{Numerical methods}}
\label{sec:num3d}
{The 3D geometrical setup corresponds to a cubic domain of side $L$, in which a planar uniform flow [in plane $(x,y)$], homogeneous in the normal direction $z$, impacts the cylindrical obstacle, with axis in the $z$ direction and height $L=16d$, thus generating a turbulent wake behind it. The 2D case is obtained by considering only the dynamics in a square domain of side $L$ [in plane $(x,y)$]. 
In the following we will also refer to the streamwise ($x$), the cross-stream ($y$) and the spanwise ($z$) directions as the longitudinal, the transversal, and the vertical one, respectively.}

In the present study, the Reynolds number is fixed. Based on the investigation on its role performed in~\cite{jaccod2021predator}, we evaluate as a reasonable choice to select it at an intermediate value, $Re=2000$, also considering the substantial increase of the computational cost of 3D direct numerical simulations (DNS), with respect to their 2D counterpart. The Schmidt number, due to numerical constraints, is also fixed, and takes the value $Sc = 1$. Consequently, the smallest relevant scale, i.e. the Batchelor scale, $\ell_B = \ell_{\nu} Sc^{-1/2}$, coincides with the viscous dissipation cutoff $\ell_{\nu}$. For the 2D case, being the turbulent dynamics governed by a direct enstrophy cascade (as observed in~\cite{jaccod2021predator}), the latter viscous scale can be estimated as $\ell^{2D}_{\nu} = (\nu^3/\langle\eta_{\nu}\rangle)^{1/6}$, where $\langle \eta_\nu \rangle$ is the mean enstrophy flux~\cite{boff}. For the 3D case, assuming that the flow is characterized by a direct energy cascade, the Kolmogorov scale is equal to $\ell^{3D}_{\nu} = (\nu^3/\langle\epsilon_{\nu}\rangle)^{1/4}$, where $\langle \epsilon_\nu \rangle$ is the mean kinetic energy dissipation rate~\cite{kolmogorov1941local,Frisch}.

{The dimensional and non-dimensional values we adopted for the biological model~\cite{hl} are reported in Table~\ref{tab1}.}
\begin{table}[htb]
\begin{center}
$
\begin{array}{|c|c|c|}
\hline
\text{Parameter} & \text{Value} & \text{Dimensionless value}\\
\hline
K                 & 108 \,\mu \mathrm{g~N~l}^{-1}    &  1\\
r\,(\beta)         & 0.3 \, \mathrm{day}^{-1}       &  4.285\\
R_m\,(\delta)      & 0.7 \, \mathrm{day}^{-1}       &  10\\
\kappa\,(\chi)      & 5.7 \, \mu \mathrm{g~N~l}^{-1}   &  0.053\\
\mu\,(\lambda)      & 0.0024 \, \mathrm{day}^{-1}     &  3.428\\
\gamma            & 0.01                           &0.01\\
\hline
\end{array}
$
\end{center}
\caption{Parameters used in the biological model. 
The symbols adopted for the nondimensional quantities appear in parentheses in the first column.
The values are consistent with typical oceanic ones, {those of $K$ and $\kappa$ are expressed in units of mass of nitrogen equivalent per liter}.}\label{tab1}
\end{table}

All the dynamical equations are solved through the open-source code Basilisk 
(\texttt{http://basilisk.fr}),
through an adaptive grid with maximum resolution $N=2^9$ for both the 2D and 3D cases. To perform a reasonable comparison between the two cases and to cope with the numerical constraints imposed by the 3D configuration, we first performed the 3D simulations, estimating the value of $\Delta x/\ell^{3D}_{\nu}$ (with $\Delta x$ the smallest mesh size), which resulted to be around $1.5$. Then, we performed several 2D simulations by varying the minimum and maximum 
resolutions in order to ensure that the mesh size is approximately the same, $\Delta x/\ell^{2D}_{\nu} \approx 1.5$. This implies that the fluid (and scalar) dynamics are moderately under-resolved and consequently the present results for the 2D case can be slightly different from the ones at the same Reynolds number from~\cite{jaccod2021predator}.

The adopted boundary conditions are such that inflow/outflow conditions are imposed 
on the left/right side of the domain, while free-slip conditions hold at the boundaries in the $y$-direction. {For the 3D case, at the top/bottom side periodic boundary conditions are imposed, to mimic a cylinder of ideally infinite height.}
On the obstacle we have a no-slip condition for the velocity while a no-flux condition is imposed for the two scalars, which are furthermore kept 
at the equilibrium values ($P_{eq},Z_{eq}$) at all sides of the domain.

{For the initial conditions, we fix the longitudinal advecting velocity to the uniform inflow value $u_0=1$ (in nondimensional units), while the transversal and vertical ones are zero. The scalar fields are initially set to their equilibrium values, then at a later time $t^*>0$, once the flow is in statistically stationary conditions, we let a localized patch of $P$ density 
enter the system from the left side. Its spatial distribution is of the form:
\begin{equation}
P(\bm{x}, t^*) = P_{eq} + P_a \, e^{(- ((x-x_0)^2 + (y-y_0)^2)/w^2)}, 
\label{pini}
\end{equation}
where $P_a = 0.5$ is the amplitude of the excitation, $(x_0,y_0) = (-2,0.5)$ its location and $w=0.9$ ($\simeq l_0$) its width. In the 3D case, the perturbation is introduced along the entire spanwise direction ($0 \leq z \leq L$).}

\begin{figure}[h!]
\captionsetup[subfigure]{labelformat=empty,justification=centering}
\captionsetup[figure]{justification=justified, singlelinecheck=off}
\begin{subfigure}{.5\textwidth}
\includegraphics[width=0.9\textwidth]{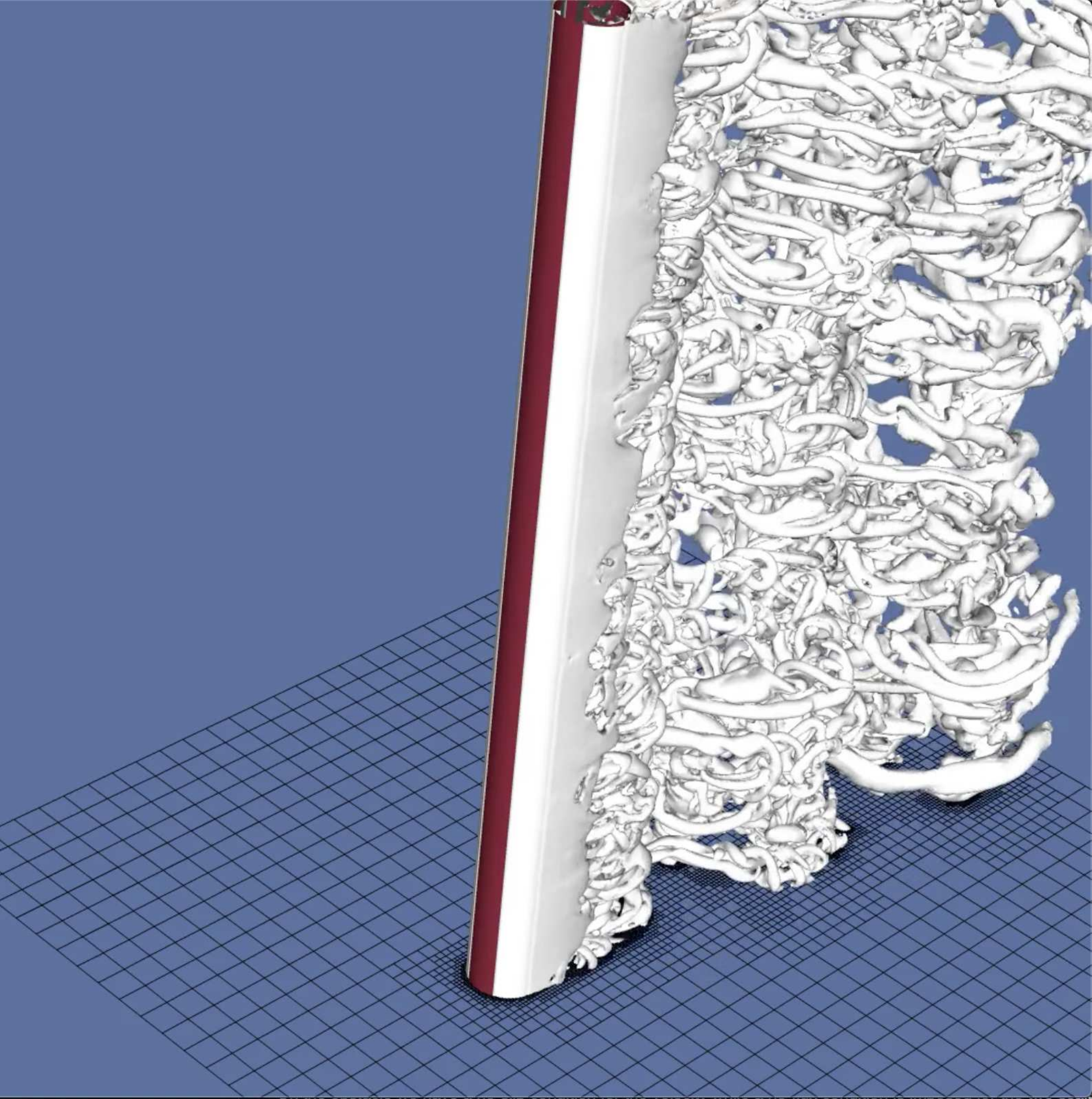}
\caption{(a)}
\label{fig:3dvort}
\end{subfigure}%
\begin{subfigure}{.5\textwidth}
\includegraphics[width=0.9\textwidth]{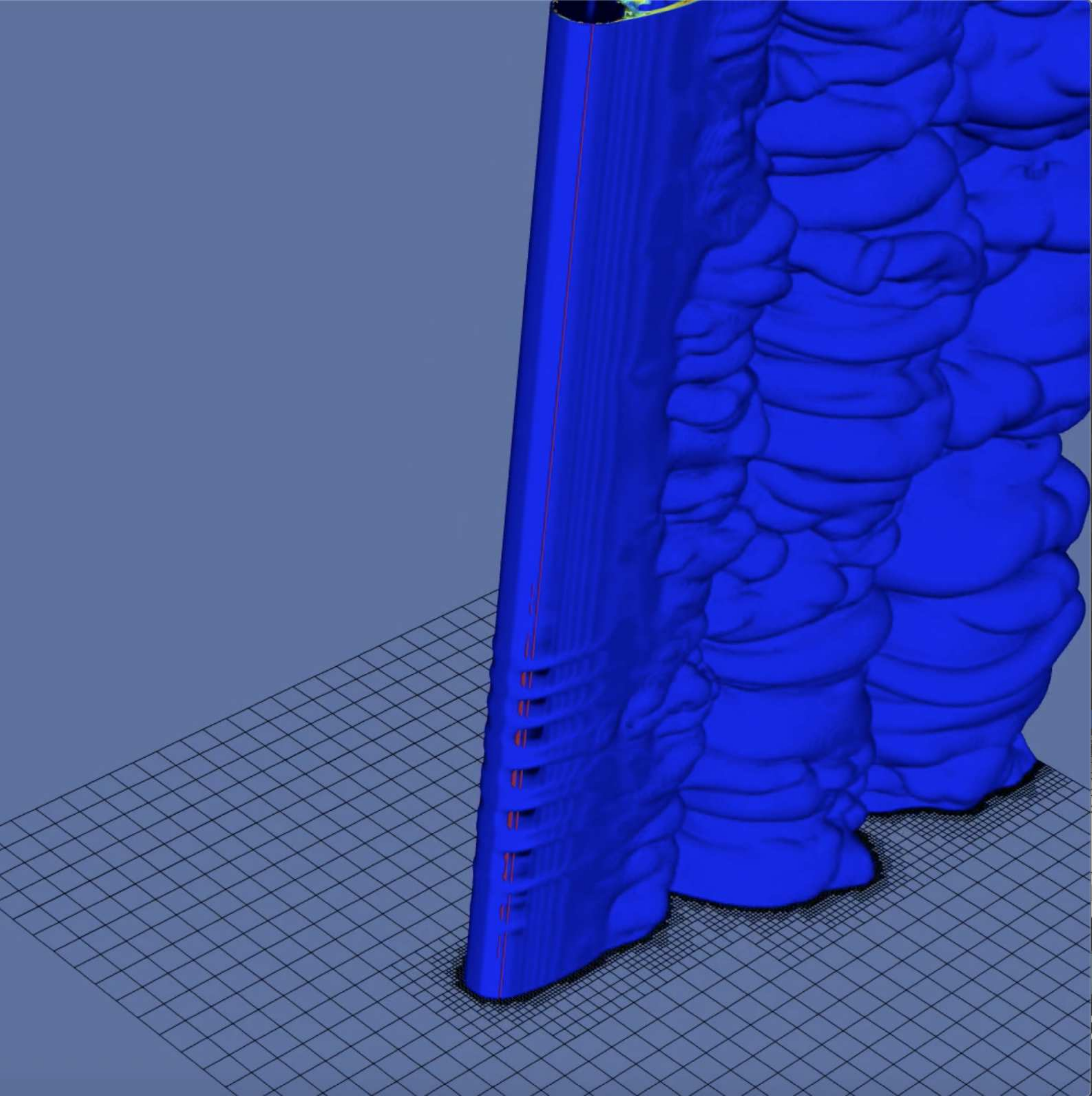}
\caption{(b)}
\label{fig:3dpl}
\end{subfigure}\\
\begin{subfigure}{.5\textwidth}
\includegraphics[width=0.9\textwidth]{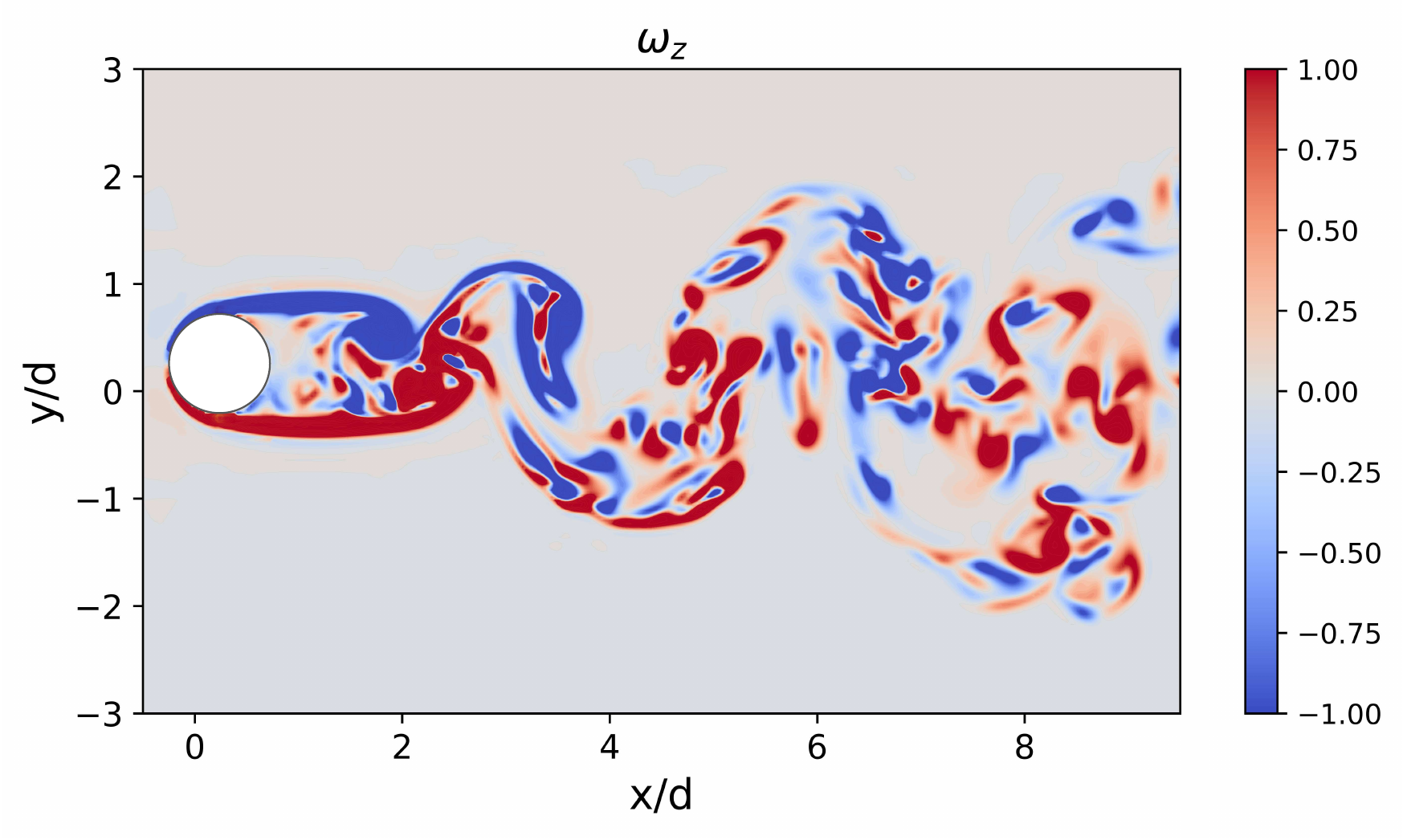}
\caption{(c)}
\label{fig:3dvort3}
\end{subfigure}%
\begin{subfigure}{.5\textwidth}
\includegraphics[width=0.9\textwidth]{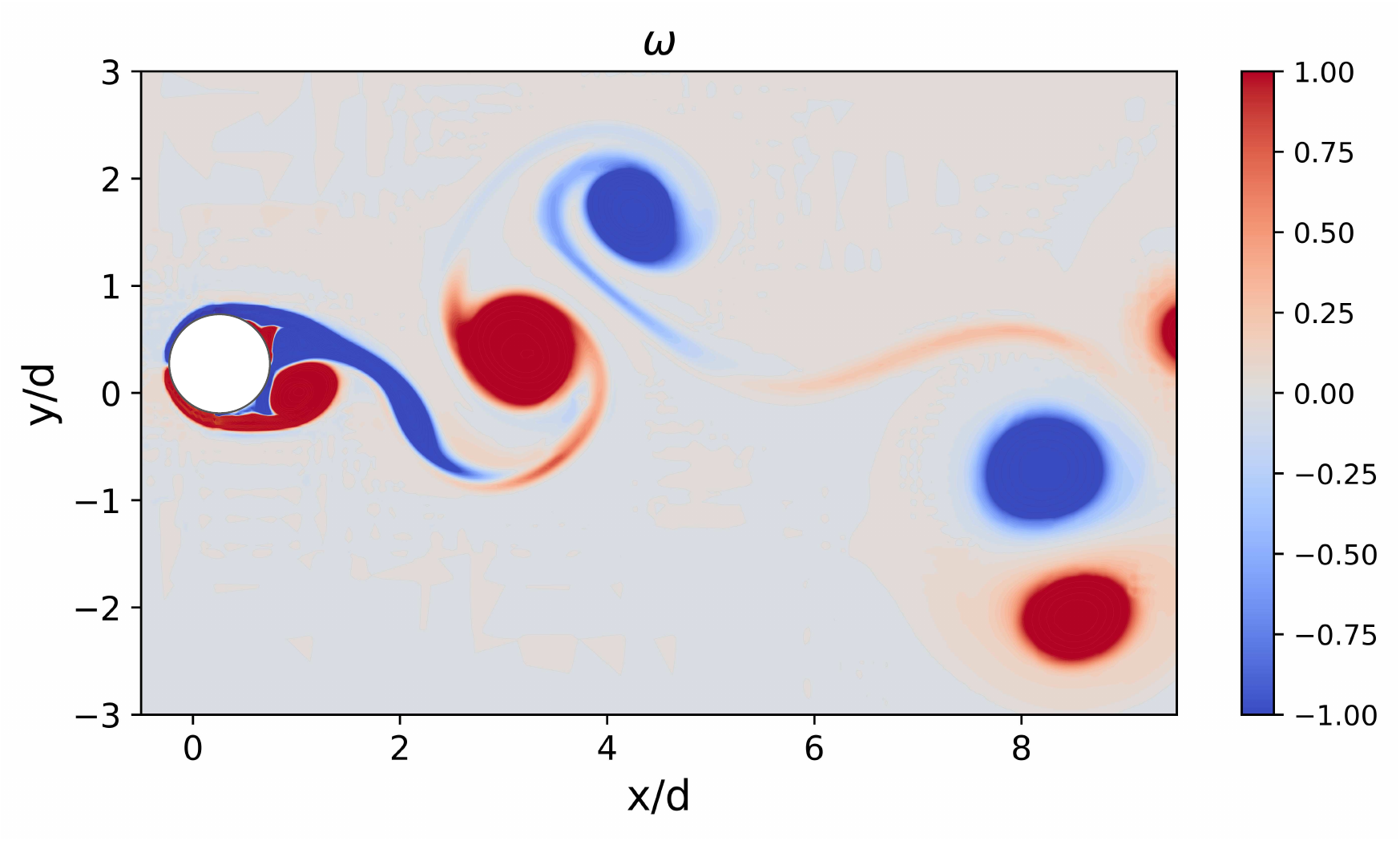}
\caption{(d)}
\label{fig:3dvort2}
\end{subfigure}
\caption{{Snapshots from the 3D simulation, at time $t=350$: (a) vorticity field (visualized through the $\lambda_2$ criterion), and (b) phytoplankton density field. The cylinder surface is shown in red, while the bottom wall displays the 
adaptive grid. 
(c) Spanwise vorticity, $\omega_z$, at $z=0$ and $t=350$ for the 3D case. (d) Vorticity, $\omega$, for the 2D case at $t=350$. To ease the comparison, in both (c) and (d), the fields have been normalized such that they take values in 
$[-1,1]$.}}
\label{fig:vis}
\end{figure}

Instantaneous visualizations of the flow and of the phytoplankton field, in 3D, are provided in Figs.~\ref{fig:vis}a and \ref{fig:vis}b. 
Here, vortices have been tracked through the $\lambda_2$ criterion~\cite{jeong_hussain_1995}. 
Figures~\ref{fig:vis}c and \ref{fig:vis}d respectively show the $z$ component of vorticity in the 3D case, and vorticity in the 2D case. 
The 3D wake displays a much better mixed {state}
with respect to the 2D one, which is characterized by coherent vortical structures with a typical size of the order of the cylinder diameter. Since the vortex-stretching term is non-zero in the 3D Navier-Stokes equation, turbulent eddies stretch along their normal direction, reducing their size until they are finally dissipated. Moreover, the 2D vortices appear to 
form much closer to the obstacle and progressively spread, covering a larger portion of the domain, with respect to the 3D case. A further characterization of the wake in the two cases will be given in the next section. 
The phytoplankton population 
distribution (Fig.~\ref{fig:3dpl}) follows the spatial organization of the flow: complex vortical structures appear immediately downstream of the obstacle and  continuously leave the domain through the right side. 


\section{Results}
\label{sec:results} 


\subsection{{Persistent blooms and global features of the dynamics}}
\label{sec:3dfullmodel}

\begin{figure}[ht]
\captionsetup[subfigure]{labelformat=empty,justification=centering}
\captionsetup[figure]{justification=justified, singlelinecheck=off}
\begin{subfigure}{.5\textwidth}
\includegraphics[width=0.9\textwidth]{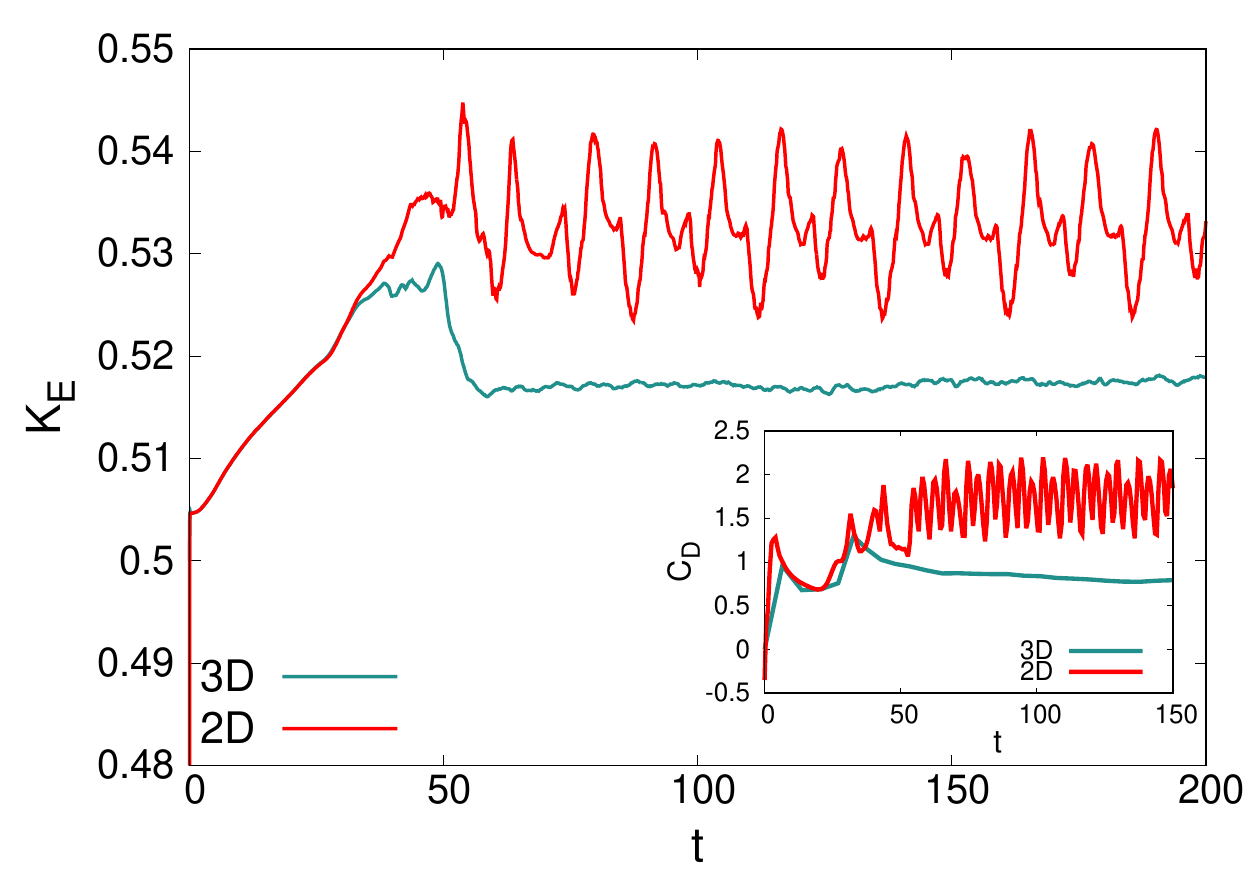}
\caption{(a)}
\label{fig:ke}
\end{subfigure}%
\begin{subfigure}{.5\textwidth}
\includegraphics[width=0.9\textwidth]{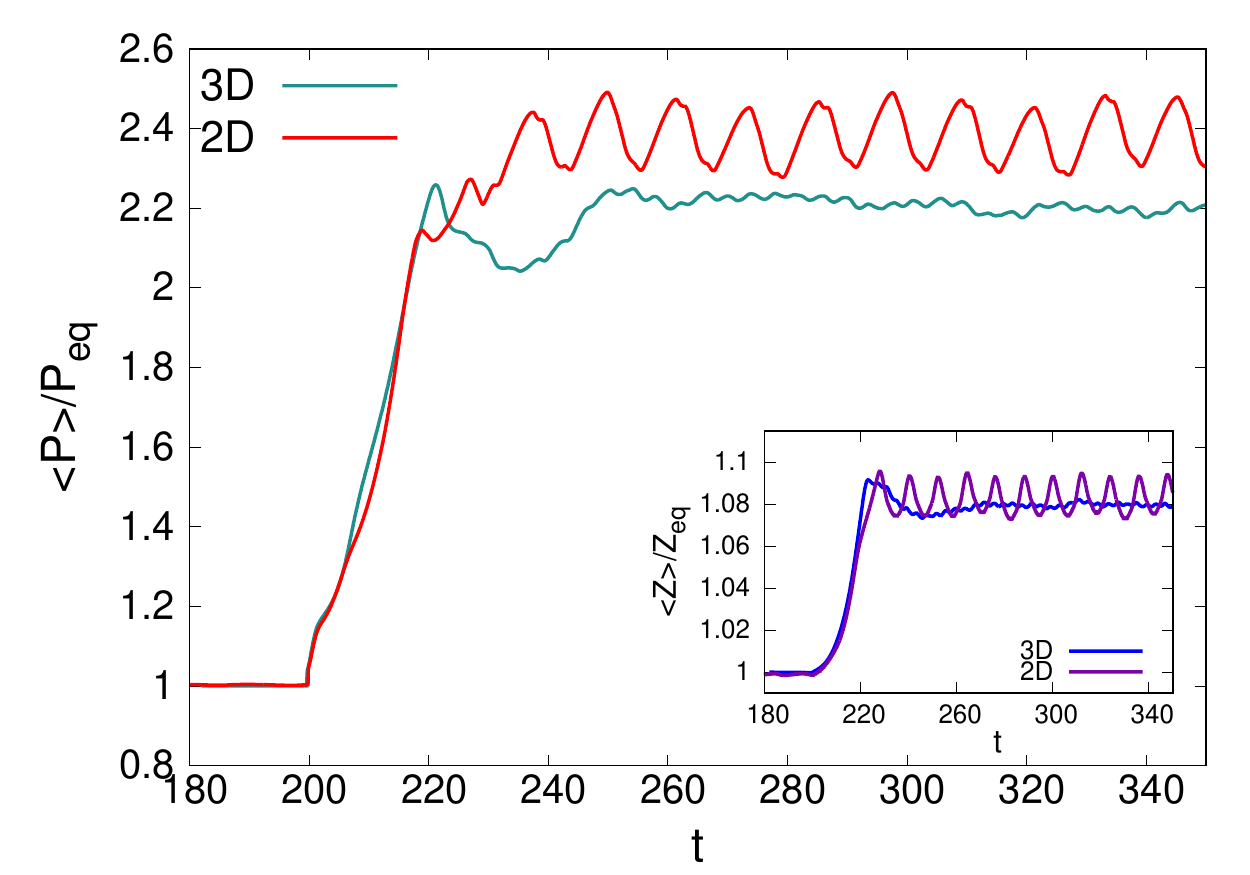}
\caption{(b)}
\label{fig:pop}
\end{subfigure}%
\caption{(a) Kinetic energy $K_E$ and drag coefficient $C_D$ (in the inset) versus time, for the 3D and 2D cases. (b) Phytoplankton population density $P$, 
 averaged in space and normalized by the corresponding equilibrium value, as a function of time for the 3D and 2D cases. In the inset we show the corresponding zooplankton population densities $Z$ vs time. 
 In panel (b) the perturbation is introduced at $t^* = 200$.
}
\end{figure}

The differences between the 3D and 2D turbulent dynamics suggested by Fig.~\ref{fig:vis} can be further appreciated in Fig.~\ref{fig:ke}. 
Here, we report the temporal behavior of kinetic energy, which shows a higher mean value and more important fluctuations in the 2D case than in the 3D one.
We also computed the integrated forces on the cylindrical solid body, in terms of the lift and drag coefficients: 

\begin{align}
C_L &= \frac{F_y}{\rho U^2 d/2}\\
C_D &= \frac{F_x}{\rho U^2 d/2}
\end{align}
where $U$ is the free-stream velocity, $F_y$ and $F_x$ are the forces per unit length in the cross-stream and streamwise directions, respectively, and $\rho$ is the fluid density.
In the inset of Fig.~\ref{fig:ke}, we show the drag coefficient $C_D$ versus time, which displays a qualitative behavior similar to that of kinetic energy. Both its mean value and its oscillation amplitudes are larger in the 2D case than in the 3D one. The same occurs for the lift coefficient $C_L$ (not shown). 
As already observed from the visualizations of Fig.~\ref{fig:vis}, coherent  structures, where vorticity is particularly intense, are created in the 2D flow. Furthermore, the roll-up of vortices happens closer to the obstacle and the near wake increases in width, hence the shedding of these structures induces larger forces on the cylinder~\cite{chua1990numerical}. In contrast, in the 3D case the roll-up occurs further downstream, so that the forces on the obstacle fluctuate less, and the width of the near wake is kept almost constant. By inferring the shedding frequency $n$ from the temporal behavior of the lift coefficient, we have calculated the Strouhal number $St = nd/U$, which resulted to be in both cases around $0.2$, in good agreement with experiments in a homogeneous non-rotating tank (where $St \approx 0.21$ \cite{z}).

The reactive dynamics in the absence of flow have been investigated in the 3D case, in analogy to what was done for the 2D case in~\cite{jaccod2021predator}. Without advection, after a sudden increase in the $P$ concentration, the effect of grazing by $Z$ makes the system come back to the equilibrium point (results not shown for the sake of brevity). 
When the advection term is switched on, the transient character of the flow, combined with the excitable character of the biological dynamics, gives rise to a permanent excitation of the predator-prey system, as also observed in kinematic and dynamic 2D settings~\cite{lo,jaccod2021predator}. 
As it can be seen in Fig.~\ref{fig:pop}, in both the 3D and 2D cases, after a transient, the spatially averaged phytoplankton population density $\langle P \rangle$ reaches  values that are considerably larger than at equilibrium. For completeness, we show in the inset of Fig.~\ref{fig:pop} also the analogous behavior for the zooplankton population density Z. 
Concerning the comparison between the 3D case and the 2D one, 
it can be noted that the 2D time-averaged value $\overline{\langle P \rangle}$ is larger than the corresponding 3D one (we will discuss this point in Sec.~\ref{sub:spa}). 
We further remark that, while the temporal behavior of the 2D case is to good extent controlled by the vortex shedding, which has a period slightly larger than $1/n$, the 3D case appears more irregular and shows oscillation amplitudes that are definitely smaller than in the 2D case. This suggests, in accordance with the behavior of kinetic energy (Fig.~\ref{fig:ke}), that at the same Reynolds number the three-dimensionality of the flow leads to more chaotic fluid motion,  
due to which the reactive scalars oscillate more irregularly in time.

Despite these quantitative differences, from a qualitative point of view, the global response of the two scalars to the combined effect of fluid transport and biological interactions seems to be the same in both the 2D and the 3D cases.
We then argue that also in the 3D case the mechanism controlling the sustained bloom (excitation of the planktonic species) is the same as the one at play in 2D systems, found in~{\cite{neu,NLHP2002,fer}} employing kinematic flows, and discussed in~\cite{jaccod2021predator} using dynamic velocity fields. 
The main ingredients are the straining action exerted by the flow, which can contrast reaction-diffusion spreading, {to localize the fast growing phytoplankton in filamentary structures along which the slowly growing zooplankton gets diluted,} and open boundaries, needed to avoid the homogenization of the scalar densities, which would end the excitation (see~\cite{NLHP2002} for more details). Moreover, the transport of biological material towards the obstacle, where strain is primarily located in the present case, is also relevant. The persistence of the excitation should then result from the characteristic timescale of strain being intermediate between the typical times of phytoplankton and zooplankton reproduction~{\cite{fer}}. 
While a 2D turbulent flow, forced at large scale, is characterized by a single timescale, determined by the strain, in the 3D case a whole range of timescales exist, associated with eddies of different sizes. 
{However, considering that large scales should still provide the largest contribution to the localization of phytoplankton, 
here we choose the timescale associated with the large-scale strain, which we dimensionally estimate as $\tau_s \sim l_0/u_0=1$.}
Using the values in Table~\ref{tab1}, we further obtain $\tau_P=\beta^{-1} \simeq 0.23$ and $\tau_Z=(\gamma \delta)^{-1} \simeq 10$ for the characteristic growth times of phytoplankton and zooplankton, respectively, and, thus, we have that the relation {$\tau_P < \tau_s < \tau_Z$} 
is satisfied. 

\subsection{{Spectral properties}}

{To characterize plankton patchiness we measured the spectra of population density variance, which we compared with the theoretical predictions from simple models of biological dynamics in turbulent flows, mentioned in Sec.~\ref{intro}, which we recall in slight more detail here below.}

{The results from idealized such idealized models vary depending on both the spatial dimensionality of the flow, and the details of the biological dynamics~\cite{DP1976,powell,franks}. If a single species is considered, in a 3D turbulent flow three regimes may be expected. In the low-wavenumber (i.e. large-scale) regime, dominated by biological growth, one should have $E^{lw}_S(k) \sim k^{-1}$, where $k$ is the wavenumber modulus. In an intermediate range of scales, dominated by turbulent motions, the plankton population density should behave like a passive (inert) scalar, and its spectrum should be $E^{iw}_S(k) \sim k^{-5/3}$.
The critical wavenumber separating these two regimes corresponds to the length scale for which the eddy turnover time and the phytoplankton growth time are comparable, $k_c = (b^3/ \epsilon_{\nu})^{1/2}$, where $b$ is a quantifier of the phytoplankton growth rate and $\epsilon_{\nu}$ is the kinetic energy dissipation rate.
At the highest wavenumbers (i.e. at the smallest scales), where a viscous-convective subrange exists only if $Sc>1$, one would have the third regime, characterized by $E^{hw}_S(k) \sim k^{-1}$. In a 2D turbulent flow, instead, the plankton variance spectrum should scale as $E_S(k) \sim k^{-1}$ at all wavenumbers~\cite{powell}. Extending these predictions to more species is not an easy task. Considering two species interacting according to a predator-prey model, namely the basic Lotka-Volterra system, it has been shown that, in the 3D case, the population density spectra of both phytoplankton and zooplankton should be given by the sum of a passive contribution $\sim k^{-5/3}$ and a reactive one $\sim k^{-3}$~\cite{powell}. In the 2D case, the same model predicts that the spectrum is not modified by biological interactions, and is then given by $E_S(k) \sim k^{-1}$ (for both populations), as for a single species.}

\begin{figure}[h!]
\captionsetup[subfigure]{labelformat=empty,justification=centering}
\captionsetup[figure]{justification=justified, singlelinecheck=off} 
\begin{subfigure}{.5\textwidth}
\includegraphics[width=0.9\textwidth]{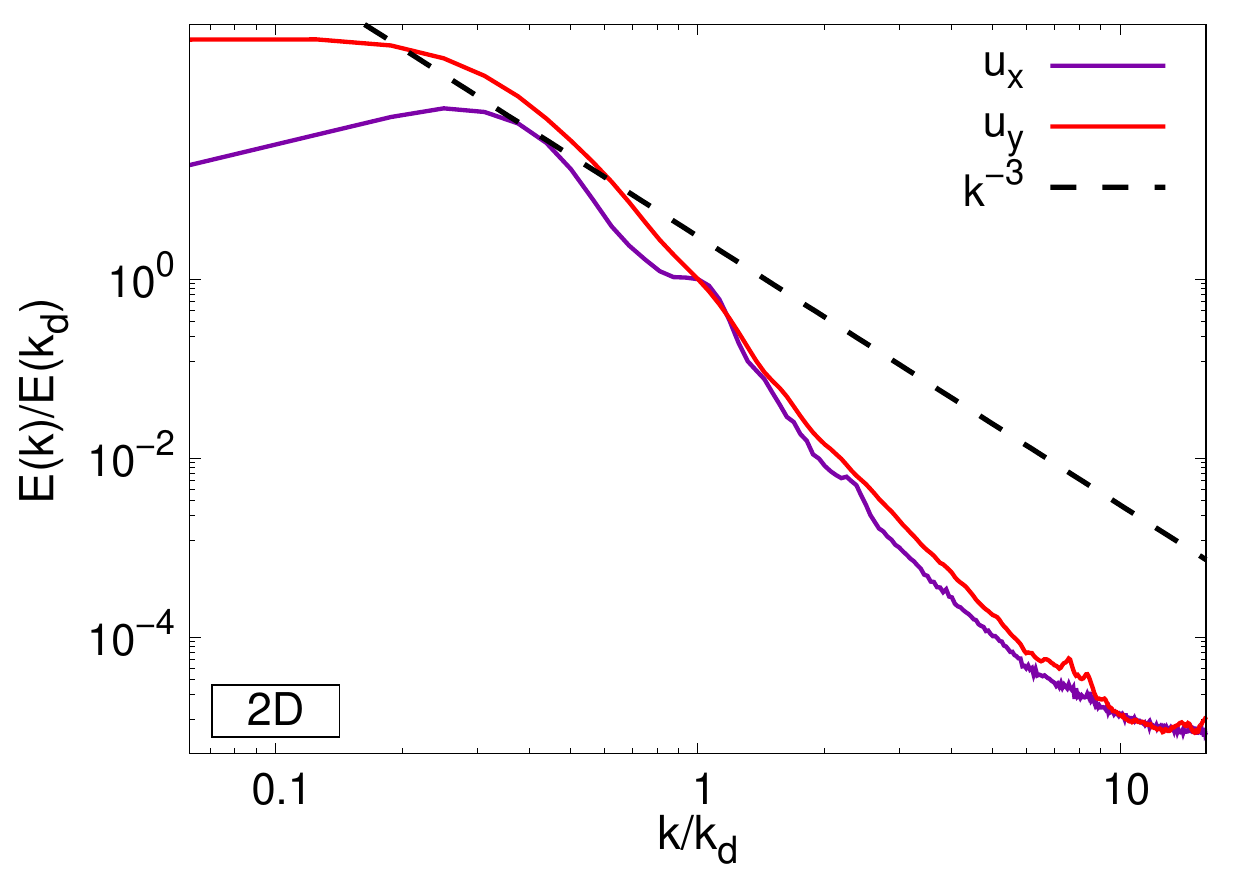}
\caption{(a)}
\label{fig:spu2d}
\end{subfigure}%
\begin{subfigure}{.5\textwidth}
\includegraphics[width=0.9\textwidth]{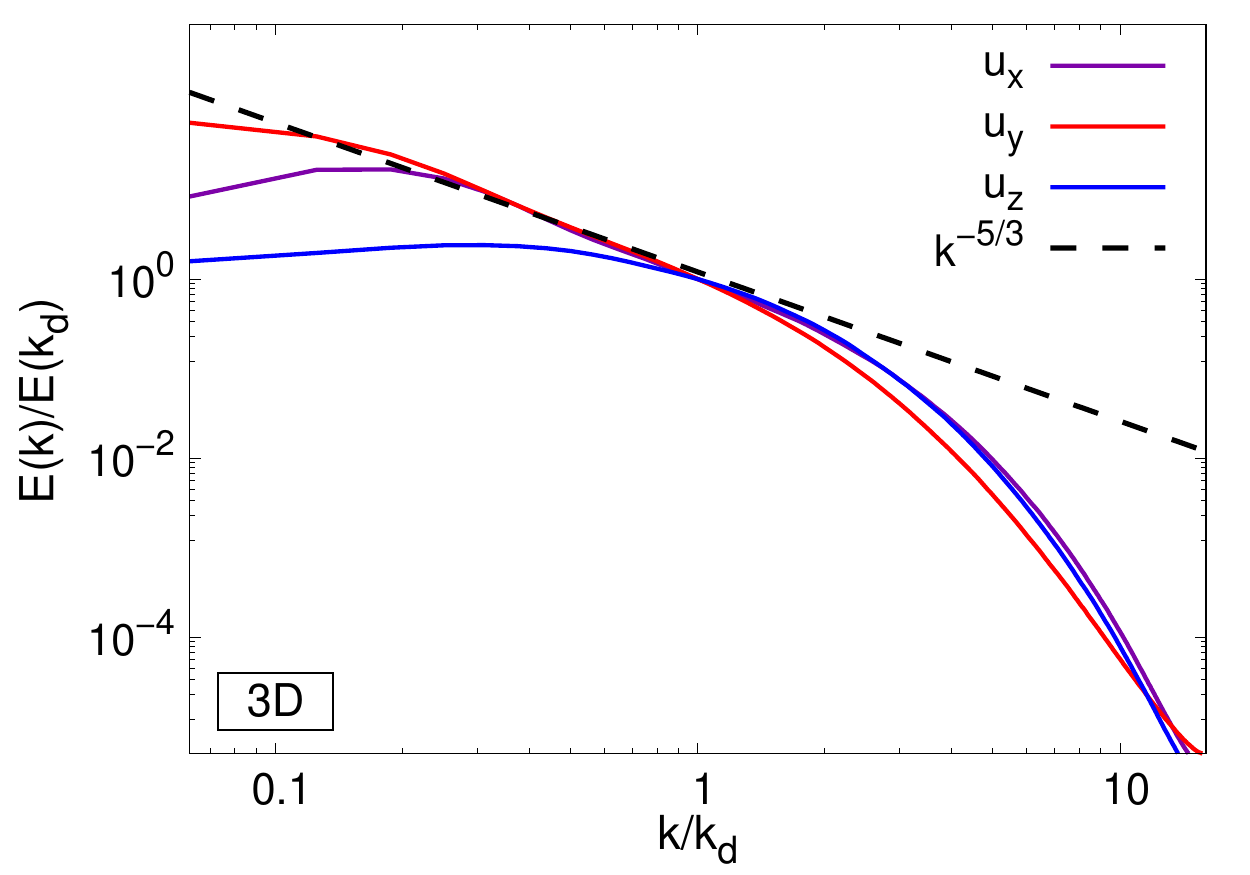}
\caption{(b)}
\label{fig:spu3d}
\end{subfigure}
\begin{subfigure}{.5\textwidth}
\includegraphics[width=0.9\textwidth]{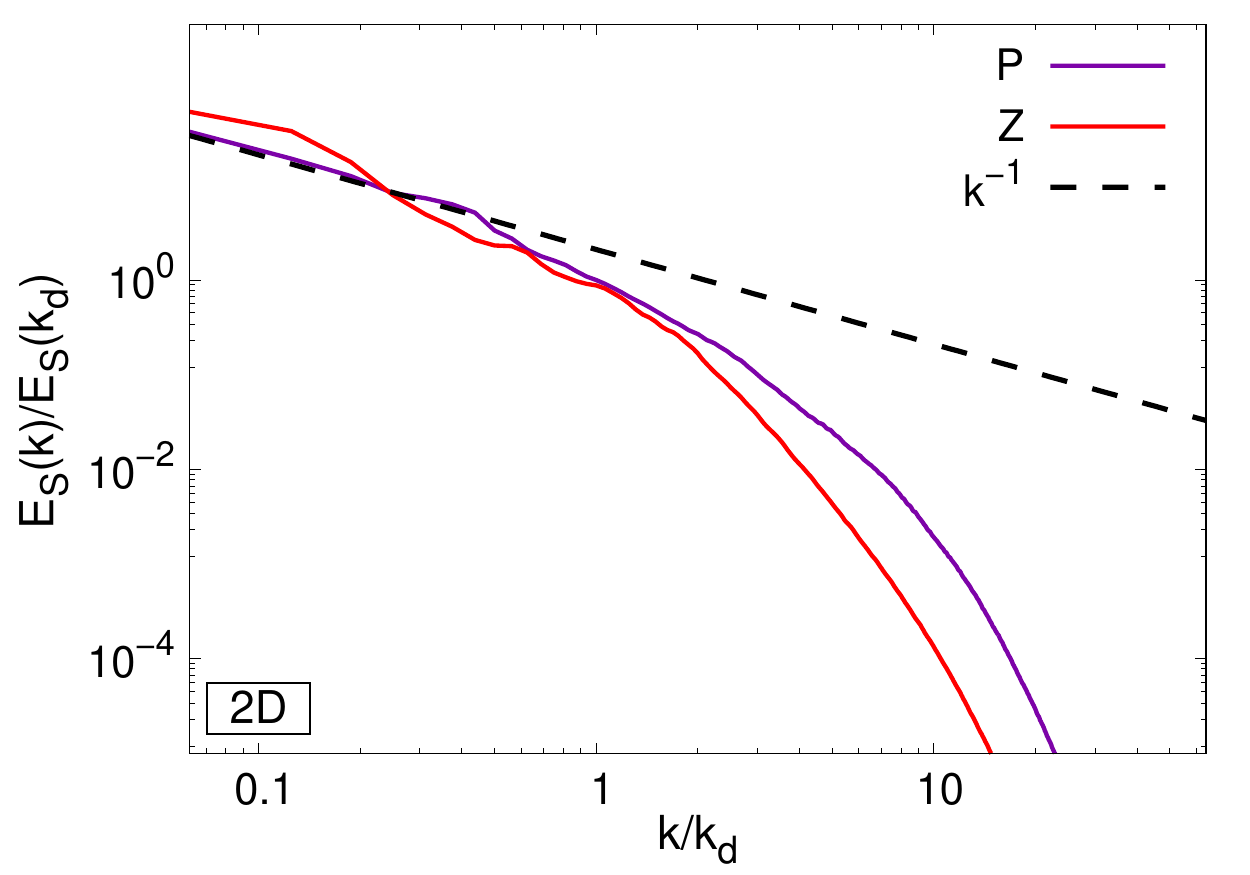}
\caption{(c)}
\label{fig:spp2d}
\end{subfigure}%
\begin{subfigure}{.5\textwidth}
\includegraphics[width=0.9\textwidth]{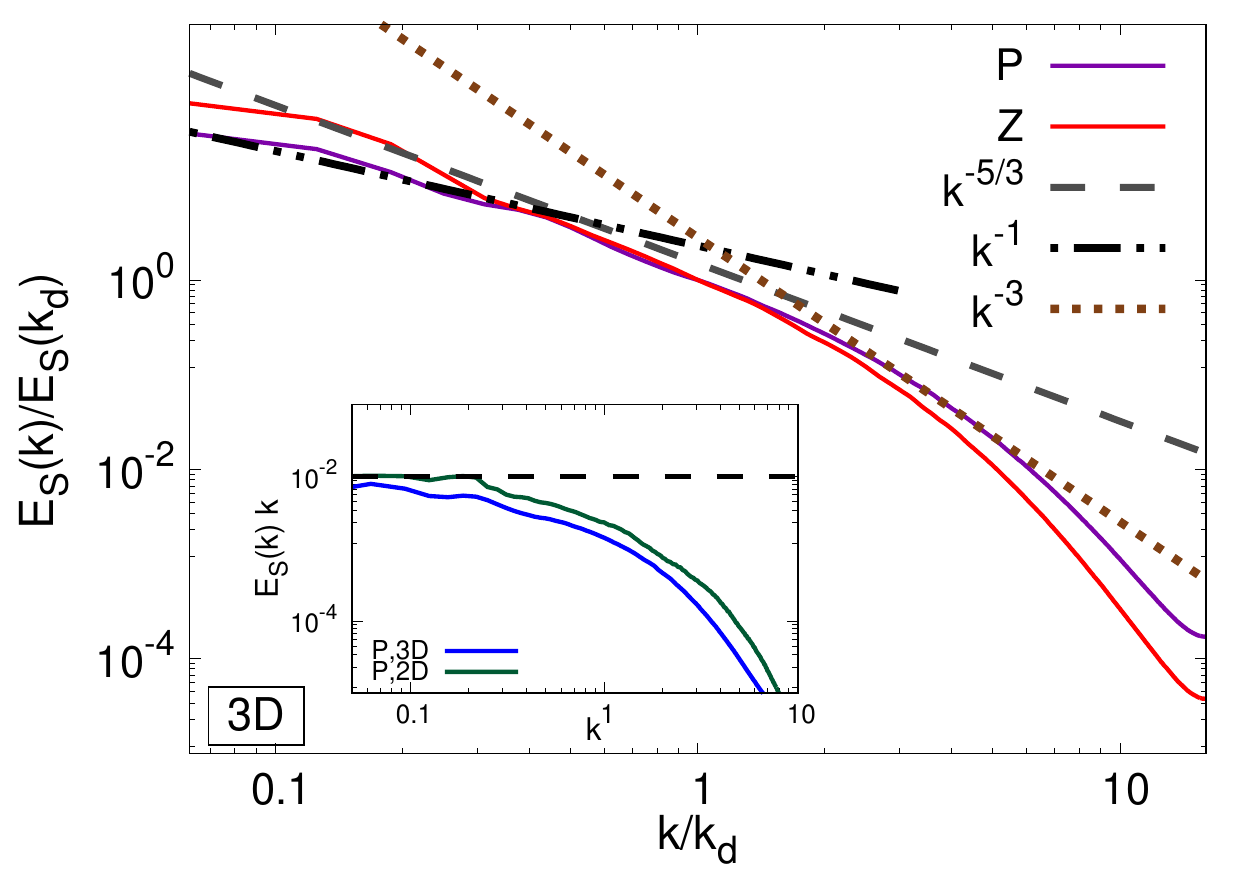}
\caption{(d)}
\label{fig:spp3d}
\end{subfigure}%
\caption{Spatial spectra of velocity component fluctuations $E(k)$ for (a) the 2D case and (b) the 3D one. 
Spatial spectra of phytoplankton and zooplankton fluctuations $E_S(k)$ (with $S=P,Z$) are shown in (c) for the 2D case and in (d) for the 3D one. All these spectra are normalized by the value corresponding to $k_d$, the wavenumber 
associated with the obstacle diameter $d$. The spectra are computed along the $y$-direction and then averaged for $1.5d \leq x \leq 10d$ and $250 \leq t \leq 400$. For the 3D case, the spectra are first computed on each plane 
$z=\mathrm{const}$ and then averaged along $z$. In the inset of (d), 
spectra of $P$, for the 2D and 3D cases, compensated with the prediction $\sim k^{-1}$, are shown.}
\label{fig:spectra}
\end{figure}

{One-dimensional spectra of velocity fluctuations, for each component $u_x,u_y,u_z$ (with $u_z=0$ in the 2D case), from our simulations, are reported in Figs.~\ref{fig:spu2d} and~\ref{fig:spu3d}. The one-dimensional spectra of the reactive scalar ($P$ and $Z$) fluctuations, in the cross-stream direction $y$, are presented in Figs.~\ref{fig:spp2d} and~\ref{fig:spp3d}. 
All spectra are computed in the subdomain $1.5d \leq x \leq 10d$ (with $d$ the obstacle diameter), taking the Fourier transform along the $y$ direction, at fixed $x$, and subsequently  
averaging along $x$. For the 3D case, the above procedure is repeated at each $z$, with further averaging in such spanwise direction. Finally, in all cases, we perform a temporal average, in the time interval $250 \leq t \leq 400$, in which the flow is statistically stationary. }

{Before discussing the spectra of the reactive scalar fields, let us illustrate the spectral properties of our turbulent flows. Kinetic energy spectra for the 2D case (Fig.~\ref{fig:spu2d}) are in agreement with previous results~\cite{jaccod2021predator}, and scale as $E(k) \sim k^{-3}$, or a steeper power law. We remark that, in such a 2D case with large-scale forcing, the precise slope of $E(k)$ is not expected to play a major role in the interplay between fluid and biological dynamics. In fact, if the spectrum is steep enough, the flow is smooth and possesses a single time scale, determined by the strain.}

{In the 3D case, the scaling of energy spectra (Fig.~\ref{fig:spu3d}) is compatible with $E(k)\sim k^{-5/3}$ over approximately one decade, pointing to the existence of a direct energy cascade~\cite{kolmogorov1941local,Frisch}. At higher wavenumbers, corresponding to unresolved scales, the spectrum rapidly falls off. We note that, at low wavenumbers, the energetic content of spanwise velocity ($u_z$) fluctuations is significantly smaller than that of the other velocity components, indicating that coherent structures arising from the cylinder three-dimensionality contain much less energy than purely 2D structures.}

{In the 2D case, the scalar fluctuation spectra (Fig.~\ref{fig:spp2d}) display a wavenumber range of power-law dependence close to $E_S(k) \sim k^{-1}$ (with $S=P,Z$), in overall agreement with the theoretical prediction, followed by a rapid decay. 
Some more comments about the extent of the scaling range are provided in Appendix~\ref{app:schmidt}, with the aim of easing the comparison of the present results with those obtained in a simulation at the same Reynolds number but at higher Schmidt number, $Sc=100$~\cite{jaccod2021predator}.}

{In the 3D case, for both planktonic species (Fig.~\ref{fig:spp3d}), we find a spectrum close to $ E_S(k)\sim k^{-5/3}$, which is the typical behavior expected for a passive non-reactive tracer advected by a 3D turbulent flow~\cite{batchelor1959small}. Note that this scaling is valid in the inertial subrange. The Batchelor regime, corresponding to scales smaller than the viscous cutoff, is here absent because $Sc=1$, i.e. $\ell_B = \ell^{3D}_{\nu}$, and scalar fluctuations are thus dissipated at the Kolmogorov scale $\ell^{3D}_{\nu}$. 
At the smallest wavenumbers ($k<0.6 \, k_d$) the spectrum of the $P$ field somehow flattens a bit. To evaluate whether this behavior could be related to the $k^{-1}$ large-scale regime, one needs to estimate the critical wavenumber $k_c$. For this purpose, we measured the effective phytoplankton growth rate $b_{eff} = \partial_t \langle P \rangle/\langle P\rangle$ (in the early growth phase, $200 \leq t \leq 220$) and the energy dissipation rate (in the statistically steady state), to obtain $k_c=(b_{eff}^3/ \epsilon_{\nu})^{1/2} \approx 0.45$, which is slightly smaller than the diameter wavenumber $k_d=0.5$. 
A clear $k^{-1}$ scaling in the range $k<k_c$ is not detectable in our spectra and we cannot safely conclude about the existence of the predicted large-scale, biologically dominated, regime~\cite{DP1976}. 
Similarly, our data do not support the $k^{-3}$ prediction, either, for the reactive contribution to the spectra of interacting species~\cite{powell}, except perhaps on narrow subranges. A possible reason for this could be the peculiar form of the PZ predator-prey model we adopted, with respect to the basic formulation of Lotka-Volterra model.} 

{It is interesting to compare the phytoplankton variance spectra in the 2D and 3D cases. These spectra are shown, after compensation with the $k^{-1}$ prediction, in the inset of Fig.~\ref{fig:spp3d}. As it can be seen, below $k_d =0.5$ the compensated 2D spectrum attains a constant value, while this is not really the case for the 3D spectrum, in line with the above discussion about the large-scale regime. Furthermore, the 2D spectrum is always larger than the 3D one, pointing to more energetic fluctuations at all scales, and thus larger variability (considering that the integral of the spectrum is the variance of the population density field), in the configuration of lower spatial dimensionality.}


\subsection{Spatial distributions}
\label{sub:spa}

\begin{figure}[t]
\captionsetup[subfigure]{labelformat=empty,justification=centering}
\captionsetup[figure]{justification=justified, singlelinecheck=off}
\begin{subfigure}{.5\textwidth}
\includegraphics[width=0.9\textwidth]{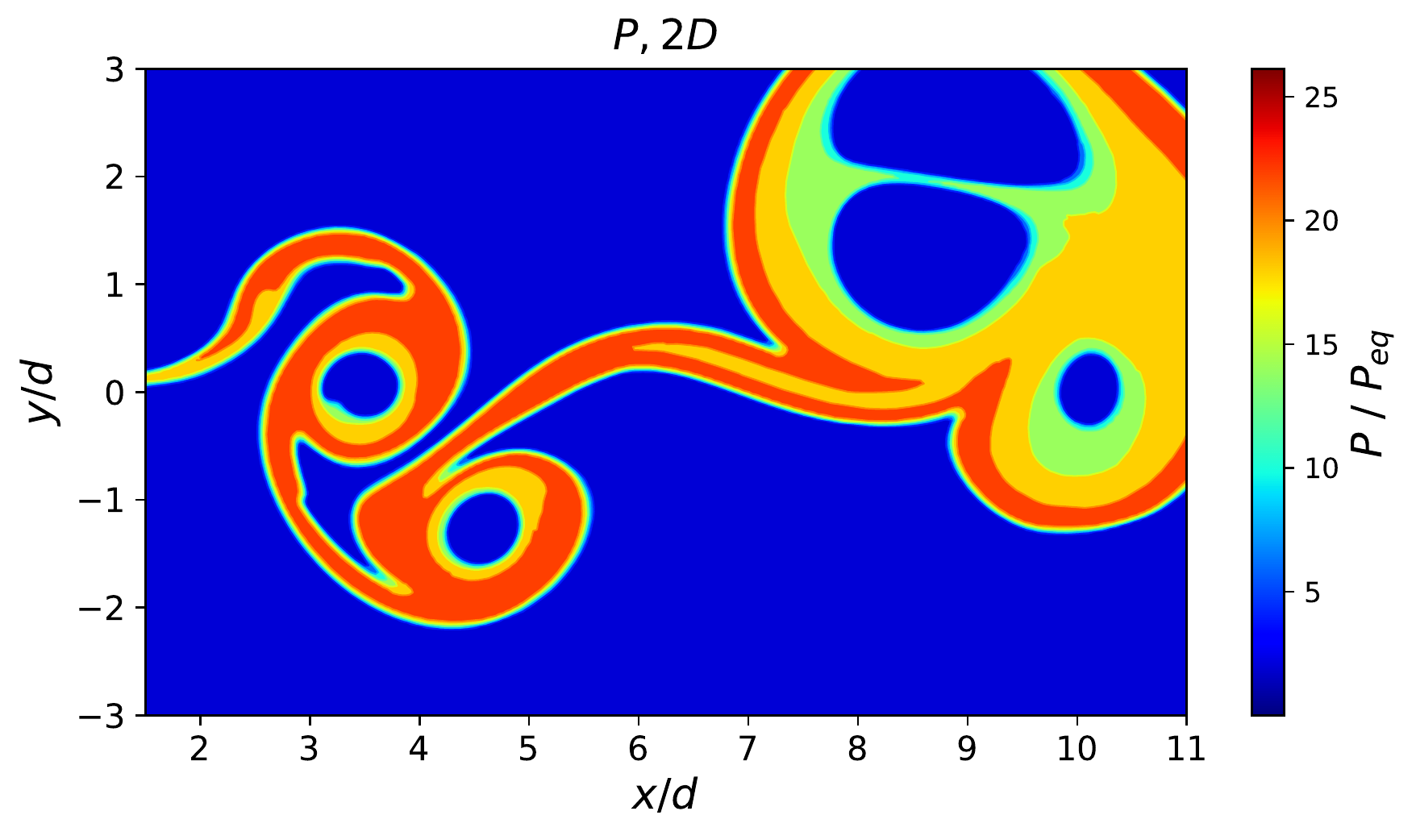}
\caption{(a)}
\label{fig:phyto2d}
\end{subfigure}%
\begin{subfigure}{.5\textwidth}
\includegraphics[width=0.9\textwidth]{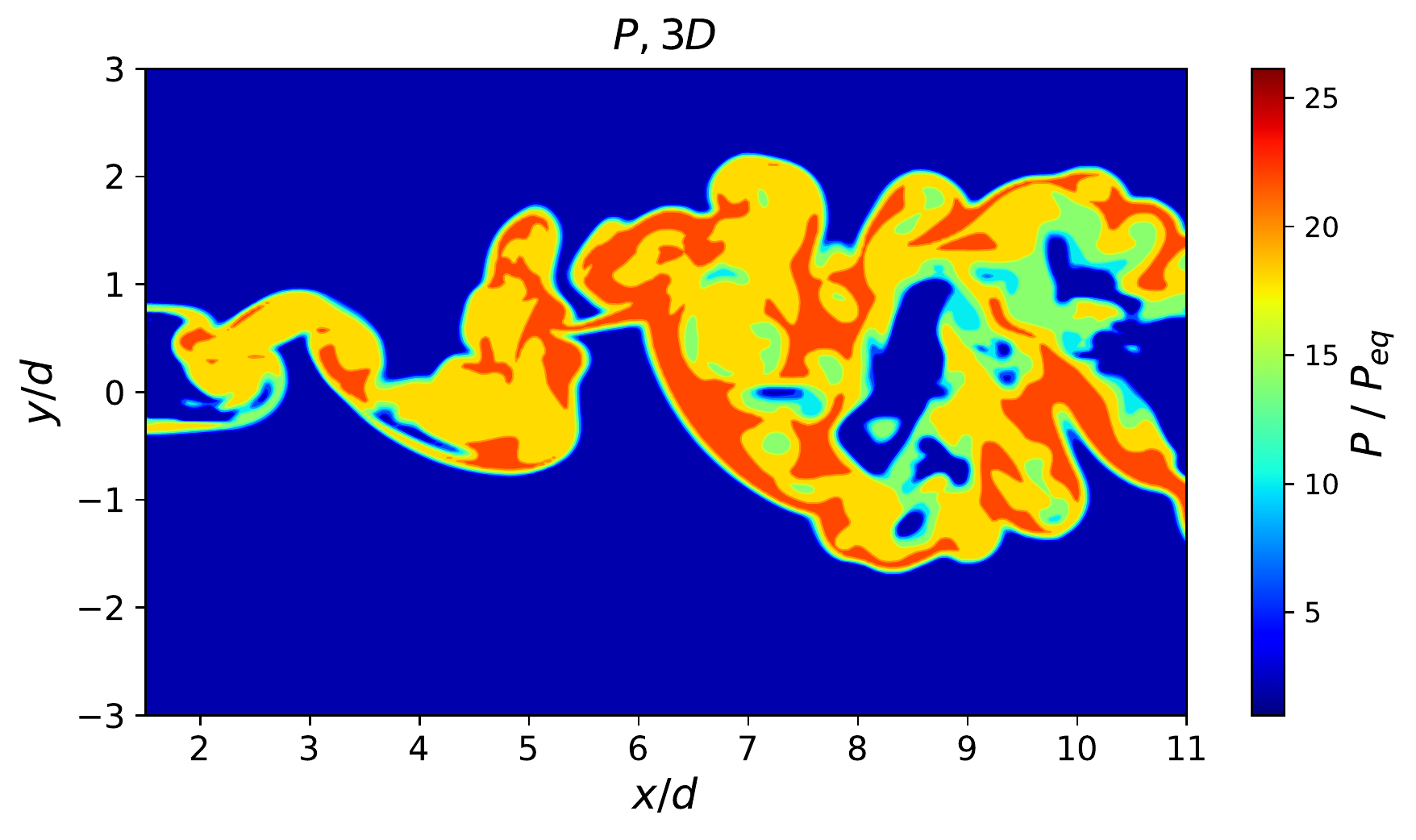}
\caption{(b)}
\label{fig:phyto3d}
\end{subfigure}
\begin{subfigure}{.5\textwidth}
\includegraphics[width=0.9\textwidth]{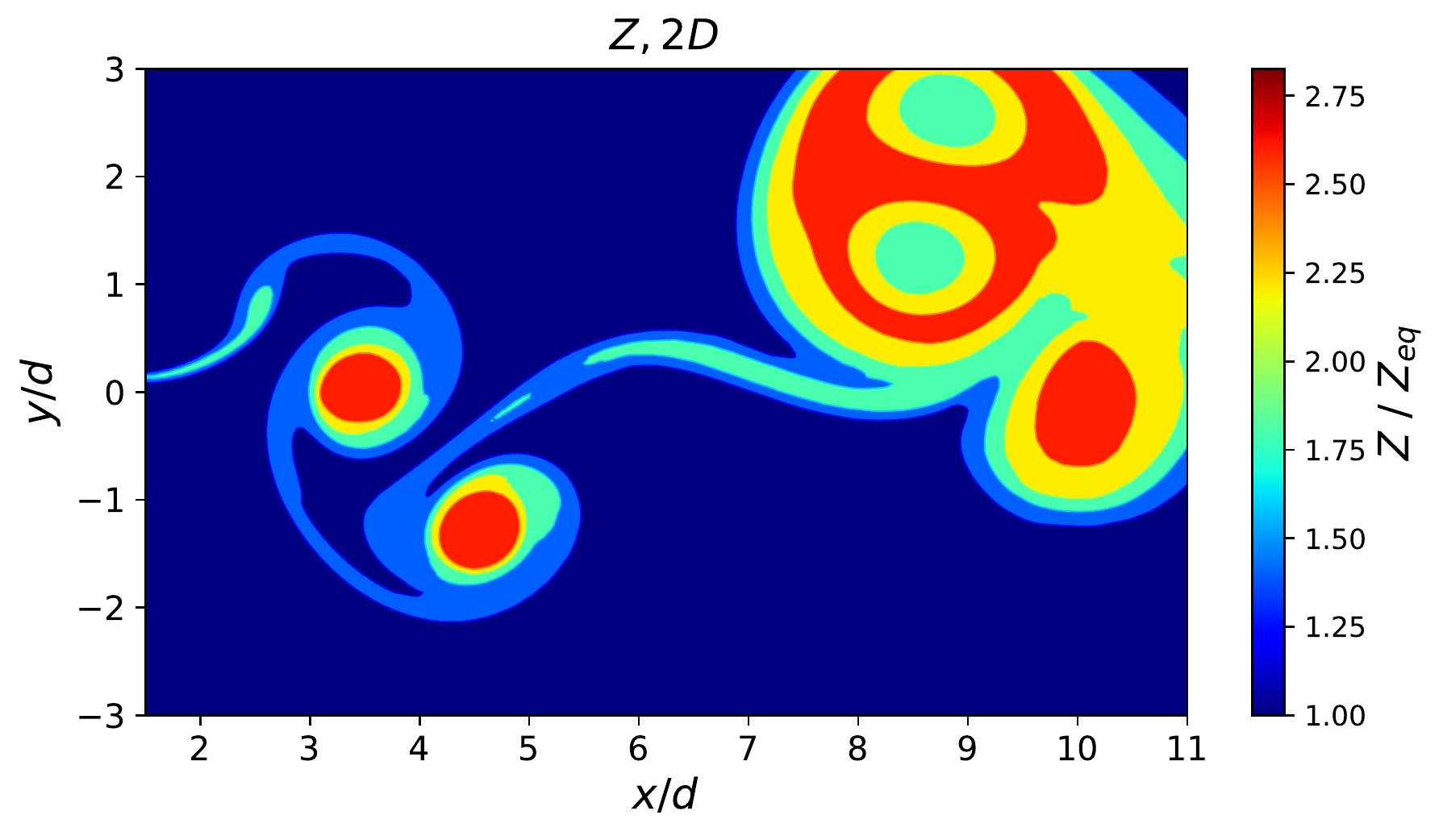}
\caption{(c)}
\label{fig:zoo2d}
\end{subfigure}%
\begin{subfigure}{.5\textwidth}
\includegraphics[width=0.9\textwidth]{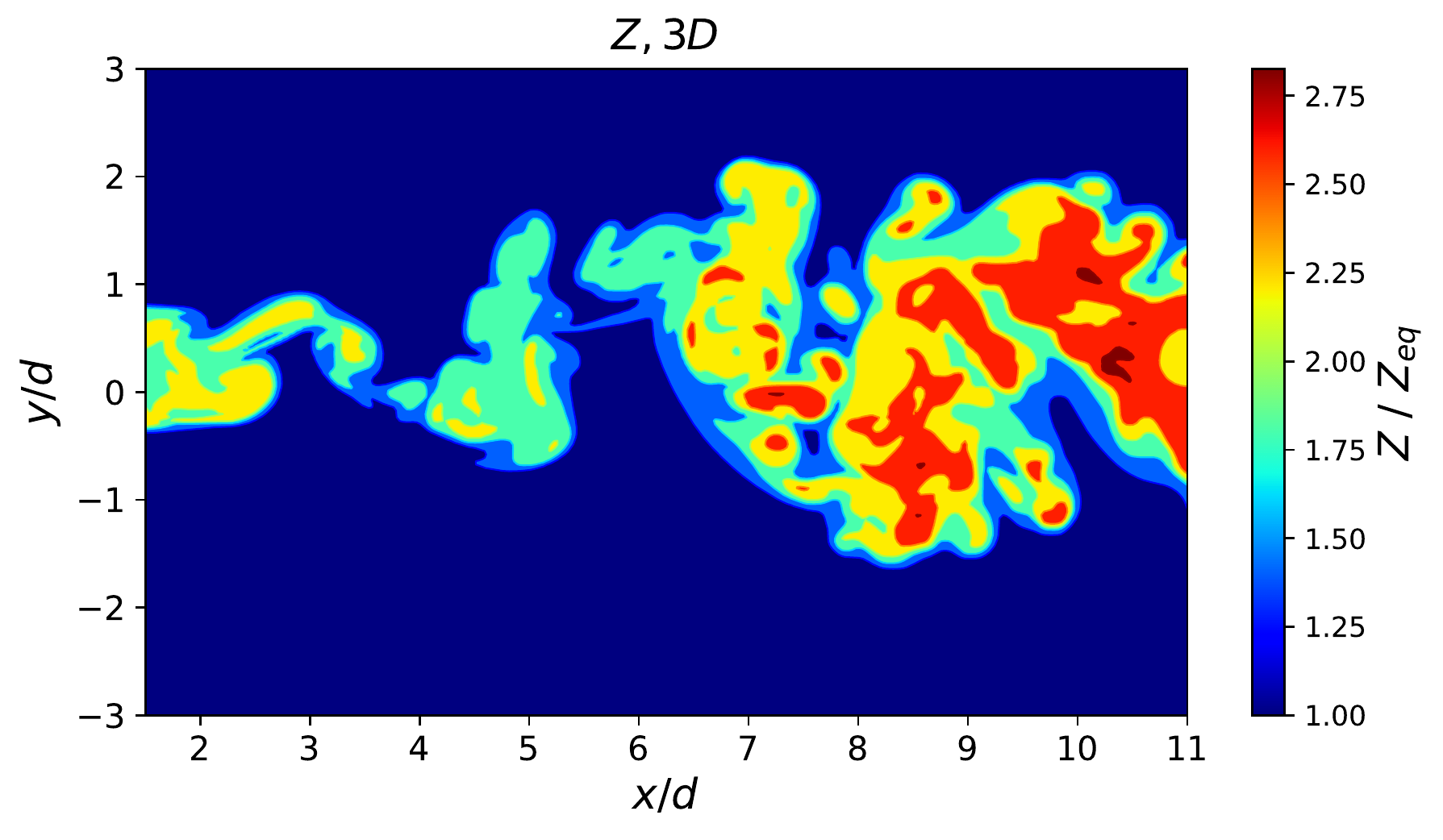}
\caption{(d)}
\label{fig:zoo3d}
\end{subfigure}%
\caption{
Instantaneous density fields of: phytoplankton $P$, panels (a) and (c) for the 2D and 3D cases, respectively; 
zooplankton $Z$, panels (b) and (d) for the 2D and 3D cases, respectively. All these fields correspond to time $t = 350$, 
in the statistically steady state.}
\label{fig:fields}
\end{figure}

{In this section, we complement the previous discussion about patchiness, by investigating the distributions of the planktonic 
species in real space. Some visualizations, at a given time in the statistically steady state, are presented in Figs.~\ref{fig:phyto2d} 
and~\ref{fig:zoo2d} for the 2D case, and in Figs.~\ref{fig:phyto3d} and~\ref{fig:zoo3d} for the 3D case. While both the 2D fields ($P$ and $Z$ ) are characterized by rather well defined coherent structures resembling those of the vorticity field (Fig.~\ref{fig:3dvort2}), the 3D distributions do not possess this feature. In the latter case, the $P$ and $Z$ 
fields reflect the more mixed nature of the 3D wake (Figs.~\ref{fig:3dvort} and~\ref{fig:3dvort3}), which is characterized by more 
convoluted vortical structures and a complex, more fragmented, spatial distribution. 
Plankton patches then appear to have smaller sizes, 
essentially independent of the distance from the obstacle, differently from the 2D case, 
where they are filaments, and almost circular structures colocating with vortices that 
grow in size and do not travel along straight lines.}  

{Nevertheless, independent of the space dimensionality, the relative abundance of the two species is locally determined 
by the predator-prey biological interactions, with the prey ($P$) mostly localized where the predator ($Z$) is absent. 
This evidence suggests that also in the 3D case, once the favorable conditions for phytoplankton growth are met, 
the response of the scalars is mainly determined by their reactive nature and not particularly sensitive to the details 
of the underlying flow.}

\begin{figure}[h!]
\captionsetup[subfigure]{labelformat=empty,justification=centering}
\captionsetup[figure]{justification=justified, singlelinecheck=off}
\begin{subfigure}{.5\textwidth}
\includegraphics[width=0.9\textwidth]{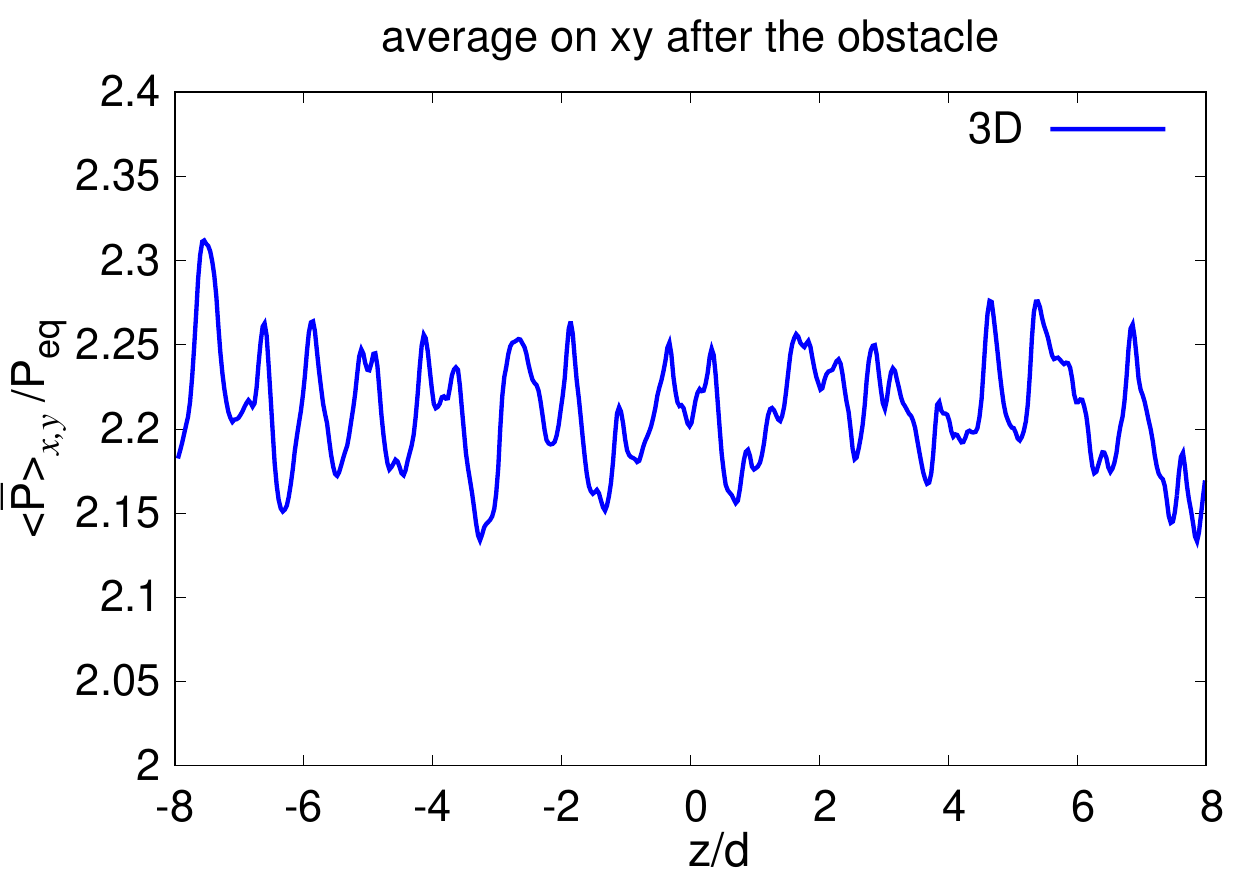}
\caption{(a)}
\label{fig:prof1}
\end{subfigure}%
\begin{subfigure}{.5\textwidth}
\includegraphics[width=0.9\textwidth]{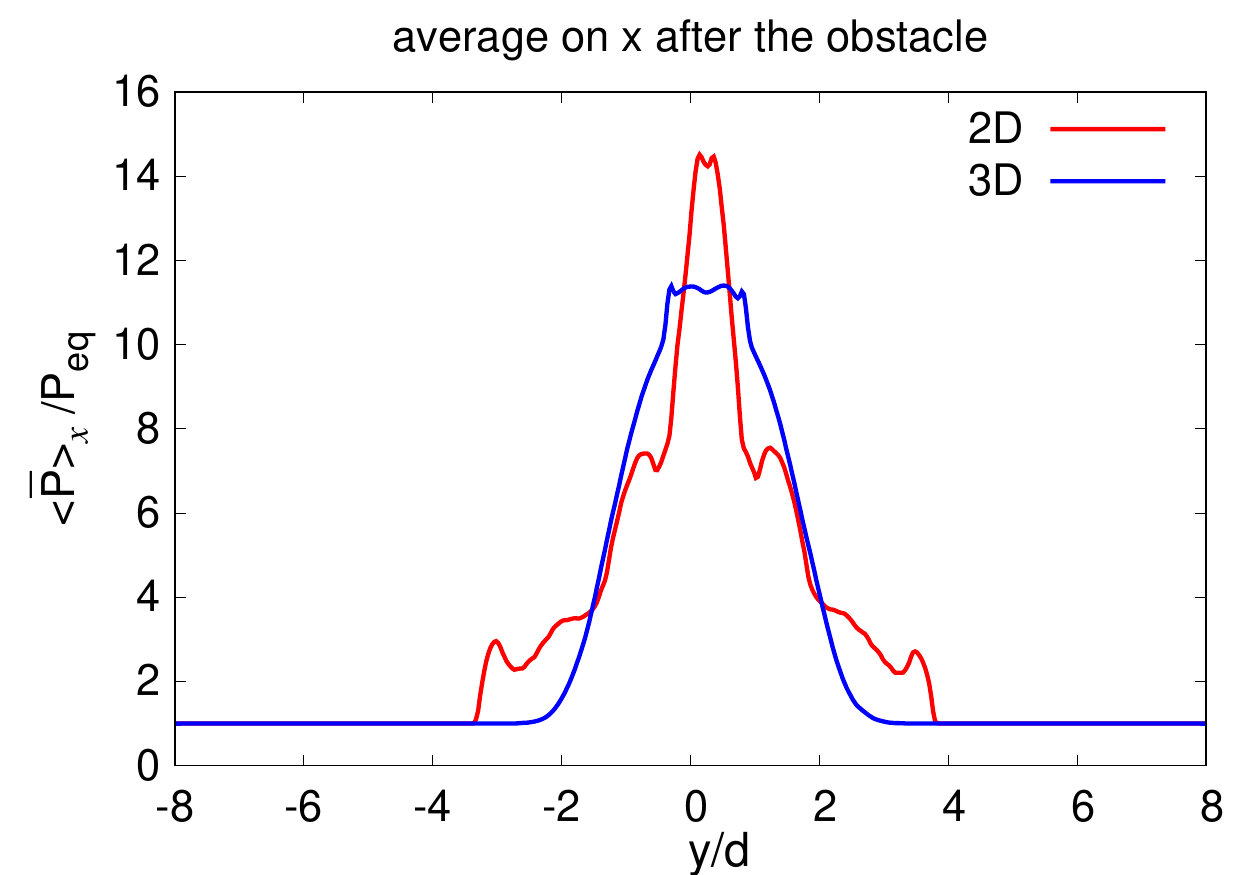}
\caption{(b)}
\label{fig:prof2}
\end{subfigure}%
\caption{
(a) Profile of phytoplankton population density in the 3D case, averaged on $x$ and $y$, 
as a function of the spanwise coordinate $z$. 
(b) Profiles of phytoplankton population density, in both the 2D and the 3D case, averaged on $x$, 
as a function of the cross-stream coordinate $y$. 
For the 3D case, the profile is averaged also on $z$. 
All the profiles shown in (a) and (b) are further averaged over time $t$ in the interval $250\leq t \leq 400$. }
\end{figure}

{We can further remark that, while the spatial distributions look different from a qualitative point of view, 
when 
passing from the 2D case to the 3D one, the range of values 
taken by each population density field does not change. 
At the same time, the spatially averaged population density is larger in the 2D case, particularly for phytoplankton 
(see Fig.~\ref{fig:pop}). To understand the origin of such difference in the global average, it is instructive to examine 
transects in the $P$ field in the two cases. 
Note that we verified that the 3D population density is, to good extent, homogeneous 
in the spanwise ($z$) direction, which allows comparing it to the 2D one. 
This is illustrated in Fig.~\ref{fig:prof1}, which shows the $P$ population density, averaged over the spatial coordinates $x$ and $y$,  
and over time, versus $z$. Indeed, such vertical profile only weakly fluctuates around a mean value ($2.2 P_{eq}$), 
which is very close to the mean value in time of the global spatial average of $P$ (Fig.~\ref{fig:pop}).}

{In Fig.~\ref{fig:prof2}, the profiles of $P$ are shown as a function of the cross-stream coordinate $y$, 
after averaging over time and $x\in[1.5d,10d]$ (and on $z$, for the 3D case). In the 3D case, phytoplankton appears 
to be predominantly concentrated in a narrow interval centered on the obstacle ($y\in [-2d,2d]$), where it is almost 
uniformly distributed, while further away in the cross-stream direction its density rapidly decays to the equilibrium value $P_{eq}$. 
Conversely, in the 2D case, the phytoplankton profile reaches values considerably above $P_{eq}$ in a larger region, hinting at the 
larger value of the global average $\overline{\langle P\rangle}$ in this case. 
Moreover, the profile now has a more complex shape, displaying several peaks of high population density, indicative of the (average) 
cross-stream location of filamentary structures.}

\begin{figure}[h!]
\centering
\includegraphics[width=0.5\textwidth]{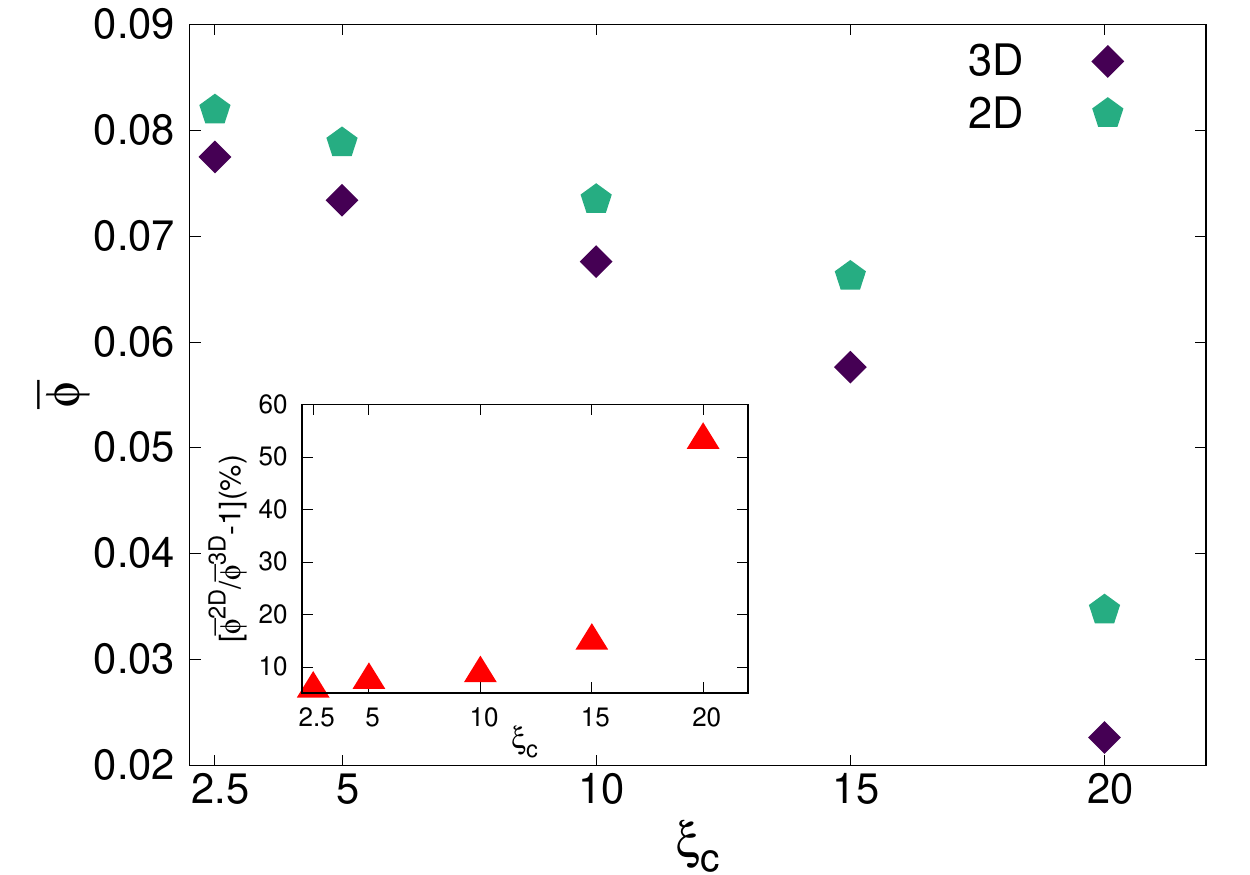}
\caption{Time-averaged fraction of the area (in the 2D case), or volume (in the 3D case), occupied by phytoplankton, 
as a function of several threshold values $\xi_c$. The inset shows the difference in percentage 
$[\overline{\phi}^{2D}/\overline{\phi}^{3D} - 1]\times 100$ vs $\xi_c$.}
\label{fig:frac}
\end{figure}

{Based on the above observations, we performed a more extensive analysis by computing the effective portion of domain 
occupied by phytoplankton. Specifically, in analogy to what is done in~\cite{berti2005mixing} for quantifying the efficiency of a reaction, 
we introduce the fraction of occupied domain, as:}
\begin{equation}
\phi = \frac{1}{\mathcal{V}}\int_{\mathcal{V}}  \theta\Biggl(\frac{P(\boldsymbol{x},t)}{P_{eq}} - \xi_c\Biggr)\ d\boldsymbol{x}
\label{eq:frac}
\end{equation}
where $\mathcal{V}$ is the domain area (in 2D) or volume (in 3D), $d\boldsymbol{x}$ represents the surface/volume element, 
$\theta(\cdot)$ is a step function and $\xi_c$ a given threshold.
{In Fig.~\ref{fig:frac}, we report the time-averaged value $\overline{\phi}$ of~(\ref{eq:frac}), 
computed in the statistically stationary state, 
for the two configurations, when varying the threshold $\xi_c$ in a wide range. In both the 2D and the 3D case, 
$\overline{\phi}$ monotonically decreases with increasing $\xi_c$. However, the 2D-case values are always larger than the 3D ones, which 
confirms that phytoplankton occupies a larger region in the lower dimensional case. Interestingly, the difference between the 
2D and 3D cases increases with growing $\xi_c$, implying that regions of particularly high population density (extreme events) 
represent a significantly larger fraction of the total extent of the domain, in the 2D case. 
This feature can be more clearly appreciated in the inset of Fig.~\ref{fig:frac}, which reports the relative difference (in percentage), 
$[\overline{\phi}^{2D}/\overline{\phi}^{3D} - 1]\times 100$; e.g., for $\xi_c >20$ the increase of the 2D case with respect to the 3D one 
reaches $55 \% $.}  

{The evidences reported in this section allow to rationalize the picture emerging from the main changes, 
in terms of biological productivity, that occur when increasing the space dimensionality. Under advection by 
the 2D turbulent flow, phytoplankton gets confined in filamentary structures, generated in straining regions, 
and winding around vortices, where it can grow thanks to its high growth rate and the initially low local population density 
of zooplankton. The more chaotic, and mixing, nature of the 3D flow, instead, hinders the formation of such structures and tends to homogenize the distributions 
of the reactive scalars. As a consequence, grazing gets everywhere more efficient, implying that regions of high 
phytoplankton density become smaller and, hence, the average biomass produced is lower.}


\section{Conclusions}
\label{sec:conclu3d}
We have investigated predator-prey plankton dynamics behind a cylindrical obstacle both in 2D and in 3D turbulent flows 
at moderate Reynolds number ($Re=2000$). Our main goal was to compare these two geometrical configurations, 
in order to test the robustness of relevant findings  
from the 2D case~\cite{jaccod2021predator}, 
about the conditions for blooming, and patchiness, 
at increasing the space dimensionality. 
{The choice of a moderate value of the Reynolds number, and of a smooth obstacle (i.e. of negligible roughness), 
was motivated by the large computational cost of 3D simulations, and previous results indicating the overall weak role 
of both $Re$ and the roughness (see~\cite{jaccod2021predator} for an extended discussion).}

{Notwithstanding the important differences between the carrying velocity fields, the qualitative behavior of the reactive scalars 
(the phytoplankton and the zooplankton population densities) appears to be similar in the 2D and the 3D cases.  
This result substantiates the general picture drawn in the 2D case, namely that the combined effect of flow transport and biological excitability is crucial to give rise to sustained plankton blooms.
Also in the 3D setup, the persistent excitation indeed appears to depend on the characteristic timescale of large-scale strain being intermediate between the timescales of phytoplankton growth 
and of zooplankton reproduction, as originally discussed using theoretical arguments and 2D kinematic flows~\cite{neu,NLHP2002,fer}, and verified in 2D dynamic simulations 
in turbulent flows~\cite{jaccod2021predator}.} 

{We then investigated patchiness by analyzing the spectra of population density fluctuations. 
The evidence of a $k^{-5/3}$ scaling (for both species) in the 3D case, and no clear hints of deviations from it of biological origin,  
which would manifest as a $k^{-1}$ or $k^{-3}$ behavior according to theoretical predictions for a single species~\cite{DP1976} 
and two interacting ones~\cite{powell}, respectively, 
suggests that the reactive scalars behave as inert ones, from the point of view of their statistical properties.} 
{This is a further similarity with the 2D case. In the latter, our numerical results agree with a $k^{-1}$ spectrum (see also~\cite{jaccod2021predator}), 
which is the expectation for a non-reactive scalar, in 2D turbulence 
forced at large scales. Nevertheless, the reactive contribution to the spectrum should also scale as $k^{-1}$ in such a case~\cite{powell}, 
which does not allow to fully ascertain whether plankton patchiness is controlled by fluid dynamics or biological ones. 
Our results, being obtained in flows of different dimensionality, appear to rather robustly indicate the prevalence of turbulent transport, in this sense.}

{The main difference between the 2D and the 3D cases reveals in the spatial distribution of the populations, which is a consequence 
of the different structure of the corresponding wakes. The 3D flow, in fact, lacks the large coherent vortices present in the 2D one, 
which are replaced by more complex thinner vortical structures. Such difference in the spatial organization of the reactive scalars has an impact 
on global properties, 
reducing the average phytoplankton density $\overline{\langle P \rangle}$ in the 3D case, which can be understood 
considering the smaller (larger in 2D) fraction of the total volume (area in 2D) occupied by phytoplankton. In essence, by destroying well-localized structures, 
like vortices and filaments in between them, where phytoplankton rapidly reproduces, the 3D fluid dynamics hamper biological growth.}


\section*{Acknowledgements}
\noindent This work was granted access to the HPC resources [MESU] of the HPCaVe centre at UPMC-Sorbonne University.

\section*{Data Availability Statement}
\noindent The datasets generated and/or analyzed during the current study are available from the corresponding author on reasonable request.


\appendix


\section{Impact of the Schmidt number in the 2D case}
\label{app:schmidt}

\begin{figure}[th]
\centering
\includegraphics[width=0.5\textwidth]{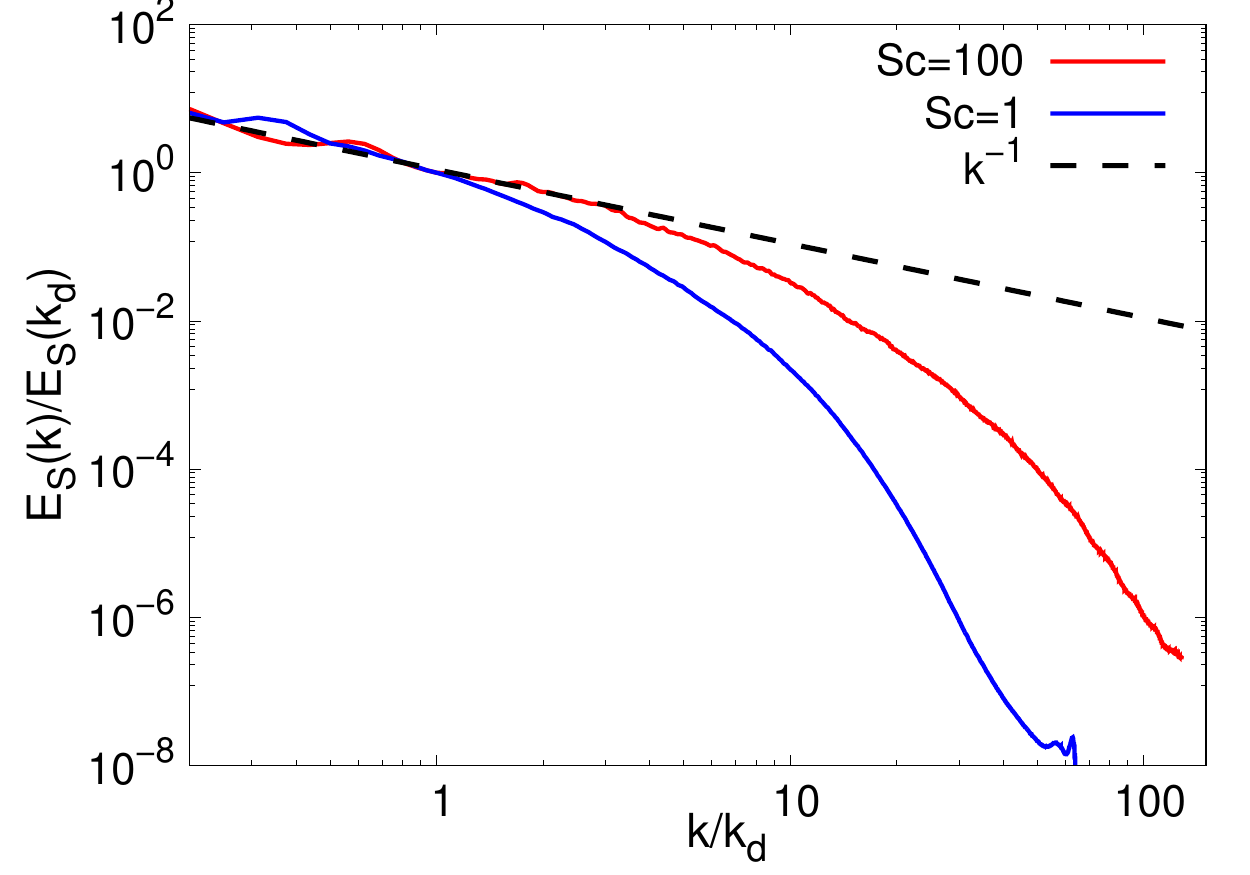}
\caption{Spatial spectra of phytoplankton density $E_S(k)$ for the 2D simulations at $Re = 2000$ and (A) $Sc=1$, (B) Sc=100,
normalized by $E_S(k_d)$, with $k_d$ the wavenumber associated with the obstacle diameter.}
\label{fig:sc}
\end{figure}

In order to 
assess the possible effect of the Schmidt number, $Sc = \nu/D$, we report here on a comparison between the results from 2D simulations performed 
at $Re = 2000$, and (A) $Sc = 1$, or (B) $Sc = 100$, focusing on scalar fluctuation spectra. 
To be consistent, in both these simulations the maximum resolution was chosen to be $N=2^{12}$, 
in order to resolve 
all the scales down to the Batchelor one, $\ell_B$.
As shown in Fig.~\ref{fig:sc}, for both (A) and (B), the wavenumber phytoplankton variance spectrum 
is compatible with the scaling $\sim k^{-1}$, but in case (A) (i.e. for $Sc=1$) this behavior is detectable only over a decade, 
for wavenumbers close to, or smaller than, $k_d$, before a rapid decay of the spectrum at larger wavenumbers. 
In case (B) (i.e. for $Sc=100$) the power-law range extends over about two decades, 
indicating higher variance at large $k$, i.e. small scales, than in (A).
This is clearly an effect due to the value of $Sc$. Indeed, when $Sc >1$, scalar fluctuations are not dissipated at the 
viscous dissipative scale but are transferred to smaller ones and finally dissipated at scale $\ell_B = \ell_{\nu} Sc^{-1/2}<\ell_\nu$. 
Considering that for simulation (B) $\ell_B \sim 0.1\ell_{\nu}$, this simple estimate explains the additional 
decade of scaling range present at $Sc=100$.

\begin{figure}[h!]
\captionsetup[subfigure]{labelformat=empty,justification=centering}
\captionsetup[figure]{justification=justified, singlelinecheck=off}
\begin{subfigure}{.5\textwidth}
\includegraphics[width=0.9\textwidth]{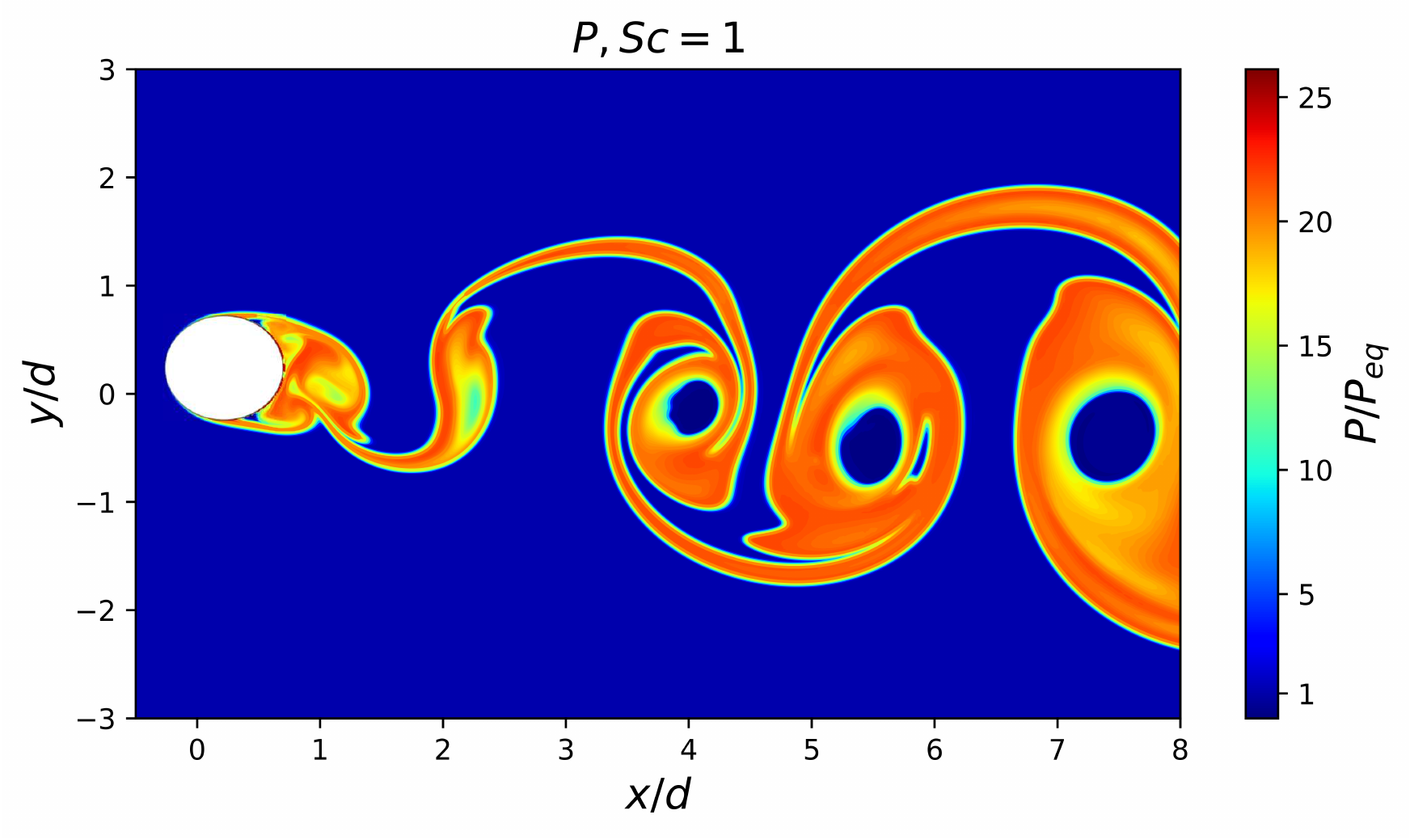}
\caption{(a)}
\label{fig:Psca}
\end{subfigure}%
\begin{subfigure}{.5\textwidth}
\includegraphics[width=0.9\textwidth]{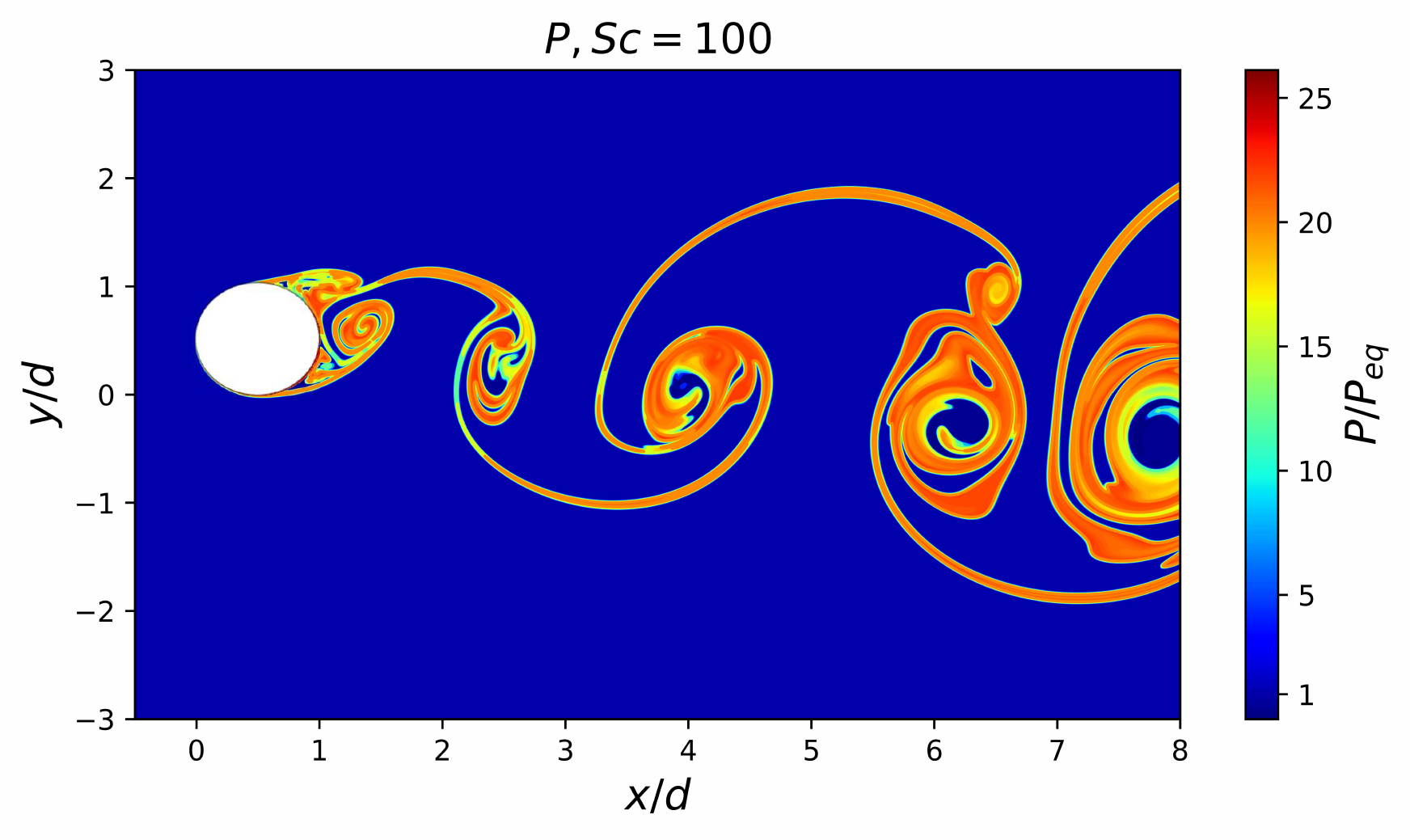}
\caption{(b)}
\label{fig:Pscb}
\end{subfigure}%
\caption{Visualization of the phytoplanklton density field at a given time for $Re=2000$ and (a) $Sc = 1$, (b) $Sc=100$.}
\label{fig:Psc}
\end{figure}

The impact of the Schmidt number can be appraised also in physical space, by comparing visualizations of the phytoplankton density field 
in the two cases (Fig.~\ref{fig:Psc}). 
For this purpose, one can consider the estimate for the width of a plankton filament in a purely elongational flow. 
According to the model originally developed in~\cite{martin2000filament}, such width is $\ell_f \sim \sqrt{D/s}$, 
where $D$ is the diffusivity and $s$ the strain rate. Note that this quantity does not depend on any biological parameters. 
In spite of the additional complexity of our flow and biological model, simulations (A) and (B) display a clear qualitative 
agreement with this estimate of $\ell_f$. The reduction of $Sc$, corresponding to the increase of $D$, makes filaments 
considerably thicker, as it is apparent from the comparison of Figs.~\ref{fig:Psca} and~\ref{fig:Pscb}.

%
%
%



%


\end{document}